\documentclass[8.5pt,twoside,twocolumn]{article}
\usepackage[super,sort&compress,comma]{natbib} 
\usepackage{mhchem}
\usepackage{times}
\usepackage{sectsty}
\usepackage{balance}
\usepackage{multirow}

\usepackage{graphicx} 
\usepackage{lastpage}
\usepackage[format=plain,justification=raggedright,singlelinecheck=false,font=small,labelfont=bf,labelsep=space]{caption} 
\usepackage{fancyhdr}
\usepackage{amssymb} 
\usepackage{color}

\begin{document}

\thispagestyle{plain}
\fancypagestyle{plain}{
\fancyhead[C]{\hspace{-1cm}\includegraphics[height=20pt]{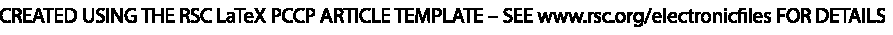}}
\renewcommand{\headrulewidth}{1pt}}
\renewcommand{\thefootnote}{\fnsymbol{footnote}}
\renewcommand\footnoterule{\vspace*{1pt}%
\hrule width 3.4in height 0.4pt \vspace*{5pt}} 
\setcounter{secnumdepth}{5}

\makeatletter 
\def\subsubsection{\@startsection{subsubsection}{3}{10pt}{-1.25ex plus -1ex minus -.1ex}{0ex plus 0ex}{\normalsize\bf}} 
\def\paragraph{\@startsection{paragraph}{4}{10pt}{-1.25ex plus -1ex minus -.1ex}{0ex plus 0ex}{\normalsize\textit}} 
\renewcommand\@biblabel[1]{#1}            
\renewcommand\@makefntext[1]%
{\noindent\makebox[0pt][r]{\@thefnmark\,}#1}
\makeatother 
\renewcommand{\figurename}{\small{Fig.}~}
\sectionfont{\large}
\subsectionfont{\normalsize} 

\fancyfoot{}
\fancyfoot[RO]{\footnotesize{\sffamily{1--\pageref{LastPage} ~\textbar  \hspace{2pt}\thepage}}}
\fancyfoot[LE]{\footnotesize{\sffamily{\thepage~\textbar\hspace{3.45cm} 1--\pageref{LastPage}}}}
\fancyhead{}
\renewcommand{\headrulewidth}{1pt} 
\renewcommand{\footrulewidth}{1pt}
\setlength{\arrayrulewidth}{1pt}
\setlength{\columnsep}{6.5mm}
\setlength\bibsep{1pt}

\twocolumn[
  \begin{@twocolumnfalse}
\noindent\LARGE{\textbf{Structures and pathways for clathrin self-assembly in the bulk and on membranes}}
\vspace{0.6cm}

\noindent\large{\textbf{Richard Matthews\textit{$^{a}$}$^{\ast}$ and Christos N. Likos\textit{$^a\ddag$}}}  \vspace{0.5cm}


\vspace{0.6cm}

\noindent \normalsize{We present a coarse-grained model of clathrin that is simple enough to be computationally tractable yet includes key observed qualitative features: a triskelion structure with excluded volume between legs; assembly of polymorphic cages in the bulk; formation of buds on a membrane. We investigate the assembly of our model using both Monte Carlo simulations and molecular dynamics with hydrodynamic interactions, in the latter employing a new membrane boundary condition. In the bulk, a range of known clathrin structures are assembled. A membrane budding pathway involving the coalescence of multiple small clusters is identified.}
\vspace{0.5cm}
 \end{@twocolumnfalse}
  ]



\footnotetext{\textit{$^{a}$~Faculty of Physics, University of Vienna, Boltzmanngasse 5, A-1090 Vienna, Austria.}}
\footnotetext{$^{\ast}$\textit{~E-mail: richard.matthews@univie.ac.at}}
\footnotetext{\textit{$\ddag$~E-mail: christos.likos@univie.ac.at}}


\section{\label{sec:intro}Introduction}

Clathrin~\cite{fotin2004} is a triskelion-shaped protein that self-assembles into a broad range of polymorphic structures. On the one hand, its key function is in forming coated vesicles, separated from membranes through budding, that are crucial for intra-cellular transport~\cite{brodsky}. On the other hand, its three-legged shape lets it also form extended hexagonal sheets~\cite{heuser}. {\it In vivo}, clathrin assembly is always associated with membranes: it is attached to them by intermediary protein complexes called adaptors.

\begin{figure}[ht!]
\begin{center}

\includegraphics[scale=0.25]{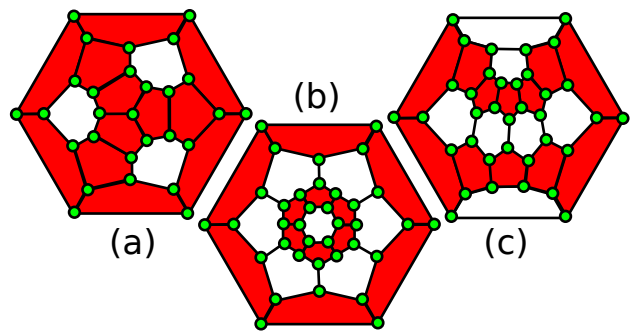}

\caption{\label{fig:Schlegel_diagrams} Schlegel diagrams of different structures assembled by clathrin, based on similar diagrams in ref.~\cite{schein2008}. (a) Mini-coat. (b) Hexagonal barrel. (c) Tennis ball. The green dots represent the centres of triskelia and the black lines joining them represent the connections between them, formed by legs lying next to each other and bonding. The red areas show pentagonal faces and the white ones show hexagonal faces. Note that the projection of the three-dimensional cage structures onto the plane does not preserve relative lengths or angles. For each diagram, the surrounding white space represents an additional hexagonal face in the three-dimensional structure.}
\end{center}
\end{figure}

Assembly of cages may also be observed in {\it in vitro} experiments without a membrane: early work~\cite{zaremba} found that the resulting cages were much more homogeneous when adaptor proteins were present. A number of closed-cage structures have been identified, all having twelve pentagonal faces and $(N - 20)/2$ hexagonal faces, where $N$ is the number of triskelia. These include one structure with $N = 28$, called a mini-coat, two with $N = 36$, given the names hexagonal barrel and tennis ball, and a truncated icosahedron with $N = 60$~\cite{pearse1987,fotin2006,schein2008}. Of these, the tennis ball structure, with a closed ring of pentagons reminiscent of the seam on a tennis ball, is less common. The mini-coat, tennis ball and hexagonal barrel structures are illustrated in Fig.~\ref{fig:Schlegel_diagrams}. Larger cages may also be formed. Detailed investigation of coated vesicles showed them to be much more poly-disperse, including some heptagons, although the tennis ball structure was also observed~\cite{cheng2007}.

Previous modeling of the assembly of structures by clathrin~\cite{otter2010,otter2010biophys} assumed that the triskelia are completely rigid. However, there is evidence, based on analysis of the fluctuations observed in electron micrograph images~\cite{jin2000} and the comparison of Brownian dynamics simulations to scattering data~\cite{ferguson2008}, that, in isolation, the legs of the triskelion have a persistence length similar to their contour length $\approx 50$nm~\cite{fotin2004}. It is however both expected~\cite{jin2000} and observed~\cite{kirchhausen2000} that there is much greater rigidity once the triskelia are bonded into a structure. 

Each triskelion leg is primarily composed of an extended, curved sub-unit called a heavy chain~\cite{fotin2004}. Much of the internal construction of the heavy chain comprises zig-zag structures. The leg flexibility within a cage was estimated, through observation of crystal structures, to allow bends of $1^{\circ} - 2^{\circ}$ per zig-zag~\cite{kirchhausen2000}. This estimate may be too high for the leg overall as it was based on a section known as the linker, which is expected to be more flexible due to a less regular structure~\cite{ybe1999}.

In this work, we present a new clathrin model that includes excluded volume. Each leg is modeled by a sequence of bonded patchy particles. The interactions between patchy particles typically have strong orientational dependence. In computational models, this may be included directly in pair interaction potentials~\cite{matthews2012} or be produced by composing sub-units of multiple particles. This latter approach may give a more realistic representation of the shape of sub-units and has been applied to viral capsids, both in studies of self-assembly~\cite{rapaport2012} and in modelling interactions with membranes without assembly~\cite{reynwar2007}, where budding was observed. Viral capsids are perhaps the most intensively studied example of self-assembly and models with single particle sub-units have also been applied~\cite{hagan2013}. For the case of clathrin, whilst previous work~\cite{otter2010,otter2010biophys} has also used such simpler models, here our approach is intermediate: sub-units composed of multiple particles whose pair interactions are patchy.

Although the exact form of the attractive interaction between clathrin legs is not known~\cite{otter2010}, in observed structures~\cite{fotin2004} they tend to lie close to each other, always having a similar relative orientation, suggesting interactions are short range and strongly orientationally dependent. Whilst, particularly for viral capsids, the use of patchy particles to represent protein-protein interactions is quite common~\cite{wilber,matthews2012,hagan2013}, we furthermore choose to employ them as an efficient way to capture the two key interaction features: short range and strong orientational dependence.

In a recent publication~\cite{matthews2012}, we presented results on the effect of fluctuating membranes on the equilibrium structures of a system of self-assembling patchy colloids. We considered a simple model, representing each clathrin with a single spherical particle with three attractive patches. Here, our use of multiple patchy beads allows, in contrast to previous approaches~\cite{otter2010,otter2010biophys}, features that are expected to be important in self-assembly to be captured: excluded volume between legs, flexibility, the interweaving of legs in assembled structures. 

The rest of the paper is organised as follows. In Sec.~\ref{sec:model}, we describe our model in more detail, including the process used to determine parameters. In Sec.~\ref{sec:bulk}, we present the results of Monte Carlo (MC)~\cite{frenkel} simulations to explore the structures that our triskelia may assemble in the bulk, before moving to dynamical simulations to consider the behaviour with a membrane in Sec.~\ref{sec:mem}. In Sec.~\ref{sec:conc}, we summarise and draw our conclusions, whereas in the Appendix we present some technical details pertaining to the model and the simulation techniques.

\section{\label{sec:model}The clathrin model}

\begin{figure}[ht!]
\begin{center}

\includegraphics[scale=0.55]{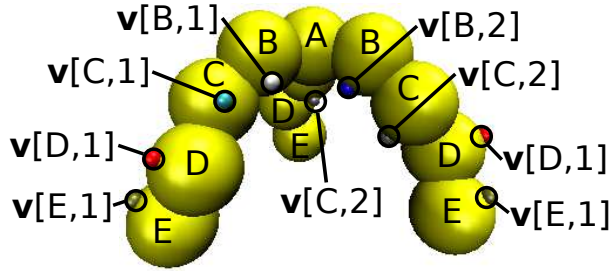}

\caption{\label{fig:triskelion_patches} Depiction of a triskelion comprised of beads type $\alpha = A - E$, with patches whose positions are defined by $\left\{\mathbf{v}[\alpha,i]\right\}$. There are attractive interactions between different triskelia, between patches  $\mathbf{v}[B,1]$ and  $\mathbf{v}[C,1]$, patches  $\mathbf{v}[B,2]$ and  $\mathbf{v}[D,1]$ and patches $\mathbf{v}[C,2]$ and  $\mathbf{v}[E,1]$. Note that all legs are identical.}
\end{center}
\end{figure}

Our model triskelion comprises 13 bead patchy beads, see appendix~\ref{app:trisk-trisk}, of 5 different types, denoted $\alpha = A - E$.  One central bead of type $A$ is attached to 3 legs, each consisting of 1 bead each of types $B-E$, see Fig.~\ref{fig:triskelion_patches}. Associated with bead type $\alpha$ are two sets of unit vectors: $\left\{\mathbf{v}[\alpha,i]\right\}$ and $\left\{\mathbf{u}[\alpha,i]\right\}$, where $i$ indexes the different vectors belonging to one type. $\left\{\mathbf{v}[\alpha,i]\right\}$ specify the attractive patches for interactions with other triskelia, whereas $\left\{\mathbf{u}[\alpha,i]\right\}$ define the internal interactions and thus the shape of an isolated triskelion in mechanical equilibrium. A further parameter, $d$, specifies the mechanical equilibrium separation between bonded beads of the same triskelion. A detailed description of how the parameters determine the triskelion shape is given in appendix~\ref{app:trisk_intern}.

Considering $\left\{\mathbf{v}[\alpha,i]\right\}$, apart from type $A$, all beads have attractive patches, depicted in Fig.~\ref{fig:triskelion_patches}, which are supplemented by a torsional vector (not shown). Moving out from the centre, type $B$ has two patches, whose positions are defined by $\mathbf{v}[B,1]$ and $\mathbf{v}[B,2]$, as does type $C$. Patches $\mathbf{v}[B,1]$ and $\mathbf{v}[C,1]$, from different triskelia, attract each other such that, if the initial parts of two legs are placed approximately antiparallel, and at an appropriate separation, they may bond. The other patches of $B$ and $C$, $\mathbf{v}[B,2]$ and $\mathbf{v}[C,2]$, attract the single patches on types $D$ and $E$, $\mathbf{v}[D,1]$ and $\mathbf{v}[E,1]$ respectively. These interactions are such that, if triskelia form a cage, the second half of a given leg bends under the first half of a leg from an adjacent triskelion, whereby ``under'' means towards the centre of the cage, mimicking nature~\cite{fotin2004}. Our choice of patches for the different beads, and of the specificity of their interactions, is made so that the parts of the legs that are observed to lie next to each other in clathrin structures~\cite{fotin2004} will attract each other in our model.

There are excluded volume interactions between all beads not belonging to the same triskelion. The shape of the triskelion is maintained by internal interactions: harmonic springs with equilibrium length $d$ and spring constant $k$ between bonded beads, plus bending and torsional stiffness. We choose the bending and torsional rigidity to be the same, specified by the parameter $\kappa$. The equilibrium angles between subsequent bonds along a leg are encoded in the $\left\{\mathbf{u}[\alpha,i]\right\}$ vectors. Full definition of the interaction potentials for both external and internal interactions is given in the appendices~\ref{app:trisk-trisk} and~\ref{app:trisk_intern} respectively. The extended nature of the legs, and the specificity of the attractive patches, gives the overall interaction between two triskelia orientational dependence. This is further enforced at a bead-bead interaction level through the torsional vectors, see appendix~\ref{app:trisk-trisk}.

The parameters of triskelion shape and patch vectors are not set {\it a priori} but they are rather specified through an informed search procedure. The basic idea of our parameter-finding scheme is the following: hold a set of objects in a desired configuration, which here will be the hexagonal barrel comprising 36 triskelia, and allow the parameters determining the interactions between these objects to vary until they have found a low energy minimum. We assume that if the interaction parameters are then fixed, they will drive the objects to reform the structure from a random initial condition. We use the term ``free assembly'' to refer to simulations where interaction parameters are fixed. For free assembly simulations we consider both random and pre-assembled initial conditions. Whilst in the latter, strong interactions may cause to triskelia to remain in their initial configuration, unlike in the parameter-finding, triskelia may in principle explore other structures.

The parameters for triskelion shape $\left\{\mathbf{u}[\alpha,i]\right\}$, patch position $\left\{\mathbf{v}[\alpha,i]\right\}$, and bead separation $d$ were chosen using Metropolis MC simulations, in which these parameters, along with the usual system coordinates, were treated as dynamical variables. Updates were made using trial moves with the standard acceptance criterion~\cite{frenkel}. Schematically, a Hamiltonian $H^{*}\left(\mathbf{X},d^{*},  \mathbf{V^{*}}, \mathbf{U^{*}} \right)$ was used to determine the parameters, which were then input to the Hamiltonian $H\left(\mathbf{X};d,  \mathbf{V}, \mathbf{U} \right)$ for free assembly simulations. $\mathbf{V}$ and $\mathbf{U}$ represent the set of all parameter vectors for all bead types and $\mathbf{X}$ represents the usual system coordinates. For $H$, the values of $d$, $\mathbf{V}$ and $\mathbf{U}$ are fixed. The corresponding variables in $H^{*}$, $d^{*}$, $\mathbf{V^{*}}$ and $\mathbf{U^{*}}$ may vary freely. Simulations with $H^{*}\left(\mathbf{X},d^{*},  \mathbf{V^{*}}, \mathbf{U^{*}} \right)$ were performed at low temperature, $k_BT \ll \epsilon_{tt}$, where $-\epsilon_{tt}$ is the minimum of the attractive interaction between beads, so that the system relaxed to a low-energy minimum. Here $k_BT$ is the energy appearing in the standard Metropolis Monte Carlo acceptance probability~\cite{frenkel}, $\min\left[1,\exp(-\Delta E/k_BT) \right]$, where $\Delta E$ is the change in the energy due to a trial move. Additional constraints were applied to ensure the minimum found corresponded to the desired structure. To extract parameter values for use in simulations with $H\left(\mathbf{X};d,  \mathbf{V}, \mathbf{U} \right)$, thermal averages of the corresponding variables around the minimum were performed. To simplify the minimisation, during the interaction-finding stage, all triskelia always had a configuration corresponding to the minimum of their internal interactions, see appendix~\ref{app:param_det} for more details.

A common self-assembled shape observed in {\it in vitro} experiments with clathrin is the hexagonal barrel~\cite{fotin2006} and we chose this as our target structure. Clearly, a different choice of target structure would lead to a somewhat different set of interaction parameters but, given its frequency in bulk assembly experiments, the hexagonal barrel is a reasonable choice. We furthermore emphasise, however, that, in free assembly runs (see Sec.~\ref{sec:bulk}), our triskelia were also able to self-assemble into different structures. Multiple parameter-finding runs at different temperatures were observed to give very similar parameters, see appendix~\ref{app:param_det}. The set of parameters used in our free assembly simulations is given in appendix~\ref{app:param_det}. The state found cannot be guaranteed to correspond to the global minimum for a hexagonal barrel. However, given the tightness of the packing of the triskelia observed in the final structure, it is a reasonable assumption that the configuration is the unique minimum for triskelia interacting in the desired way, with initial parts of legs lying side-by-side and antiparallel, and the end parts of legs tucked inside the cage. 

We set the parameter for the harmonic springs between beads, which is not varied, to $k = 1.6 \times 10^3  k_BT$, and consider different bending stiffnesses, $\kappa$ and patch attraction strengths, $\epsilon_{tt}$. Our simple model represents only those two sections of a leg, which when assembled run along two polyhedron edges. In the corresponding section of a true triskelion leg there are $\approx 20$ zig-zags~\cite{fotin2004}. We primarily consider bending stiffness parameters of $\kappa = 0.8 \times 10^3  k_BT$, $1.6 \times 10^3  k_BT$ and $3.2 \times 10^3 k_BT$. $\kappa =  0.8 \times 10^3 k_BT$ gives a typical angular deflection per bending joint of $\approx 3^{\circ}$. Since there are eight joints in a leg, one at each end of each internal bond, see appendix~\ref{app:trisk_intern}, this gives a total possible deflection per leg similar to that expected from crystal structure observations~\cite{fotin2004,kirchhausen2000}. We also considered complete rigidity, applied also to the springs joining beads, as well as flexibility similar to that seen for isolated triskelia. In the latter case we found assembly of disordered and extended structures rather than cages and results are not presented. It should be noted, however, that the structures found for stiffer triskelia will also represent local minima for the flexible ones, although in this case our free assembly simulations were unable to find them.

\section{\label{sec:bulk}Bulk self-assembly}

We next consider the structures formed by our clathrin-model without a membrane. For these free assembly simulations, we employ Metropolis Monte Carlo with a range of moves to improve sampling, including Aggregate Volume Bias~\cite{chen2001}, Configurational Bias~\cite{vlugt1998}, cluster moves~\cite{bhattacharyay2008}, Hybrid MC~\cite{mehlig1992} and multicanonical parallel tempering~\cite{faller2002}. To form closed cages, triskelia must be able to bond and form faces surrounded by both 5- and 6-edge loops. This flexibility, which is automatically built into our model through the parameter choosing procedure, means that the triskelia may explore a broad range of competing low-energy minima. Whilst we expect the global minimum to be a closed cage, simulations may easily become trapped in other states and, despite the range of MC moves utilised, we find that simulations are not able to move between all of the local minima on a feasible timescale. Nonetheless, the free assembly simulations do give us reliable information about the structures that our model may assemble.

\begin{figure*}[ht!]
\begin{center}

\includegraphics[scale=0.4]{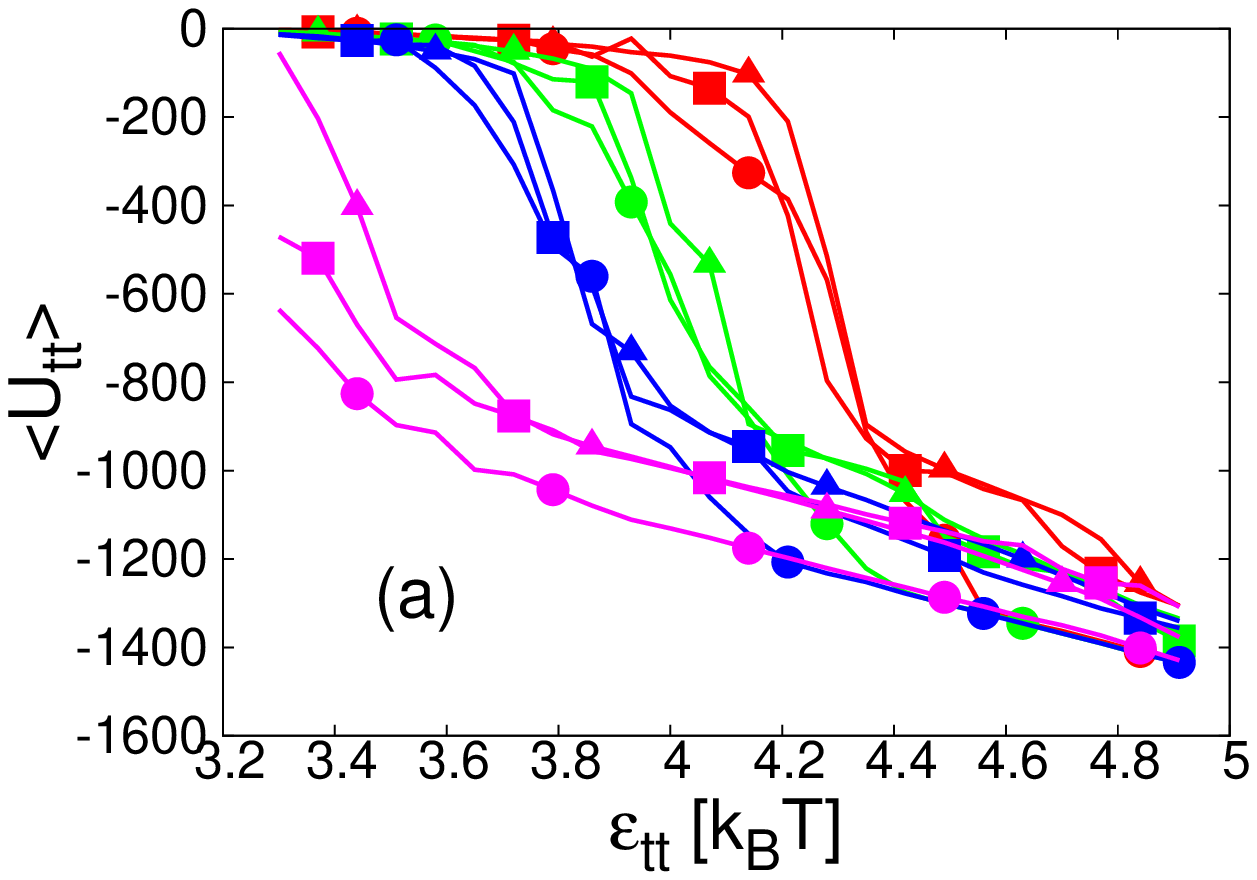}
\includegraphics[scale=0.4]{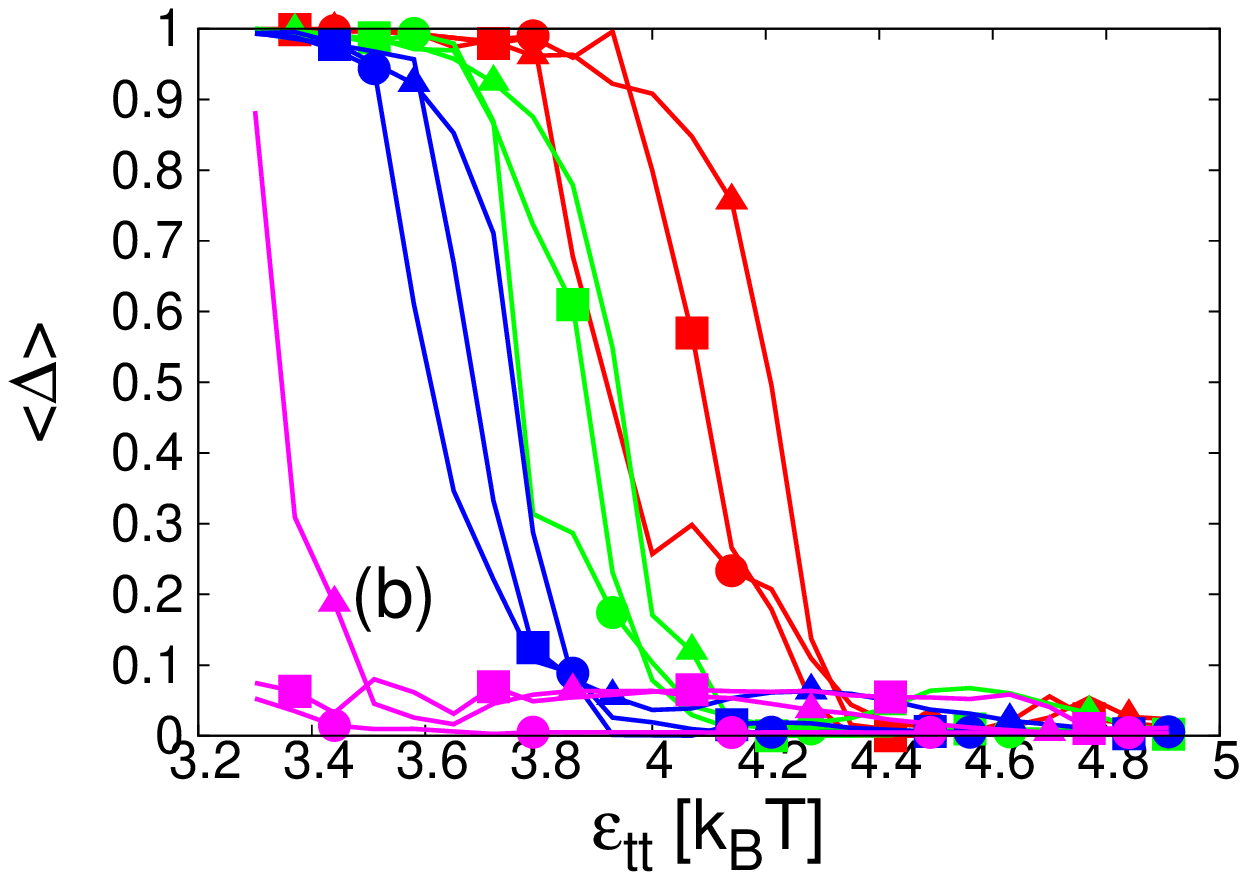}
\includegraphics[scale=0.4]{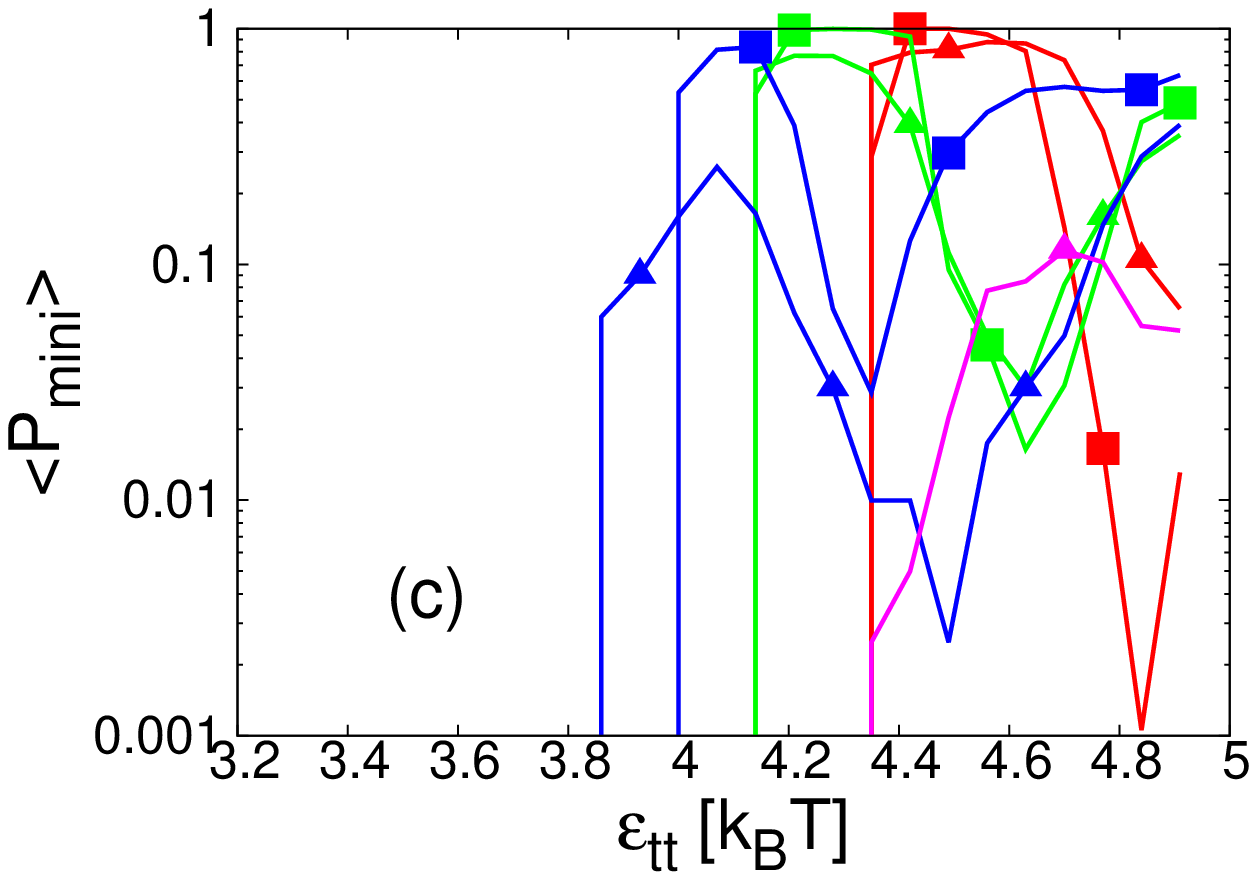}
\includegraphics[scale=0.4]{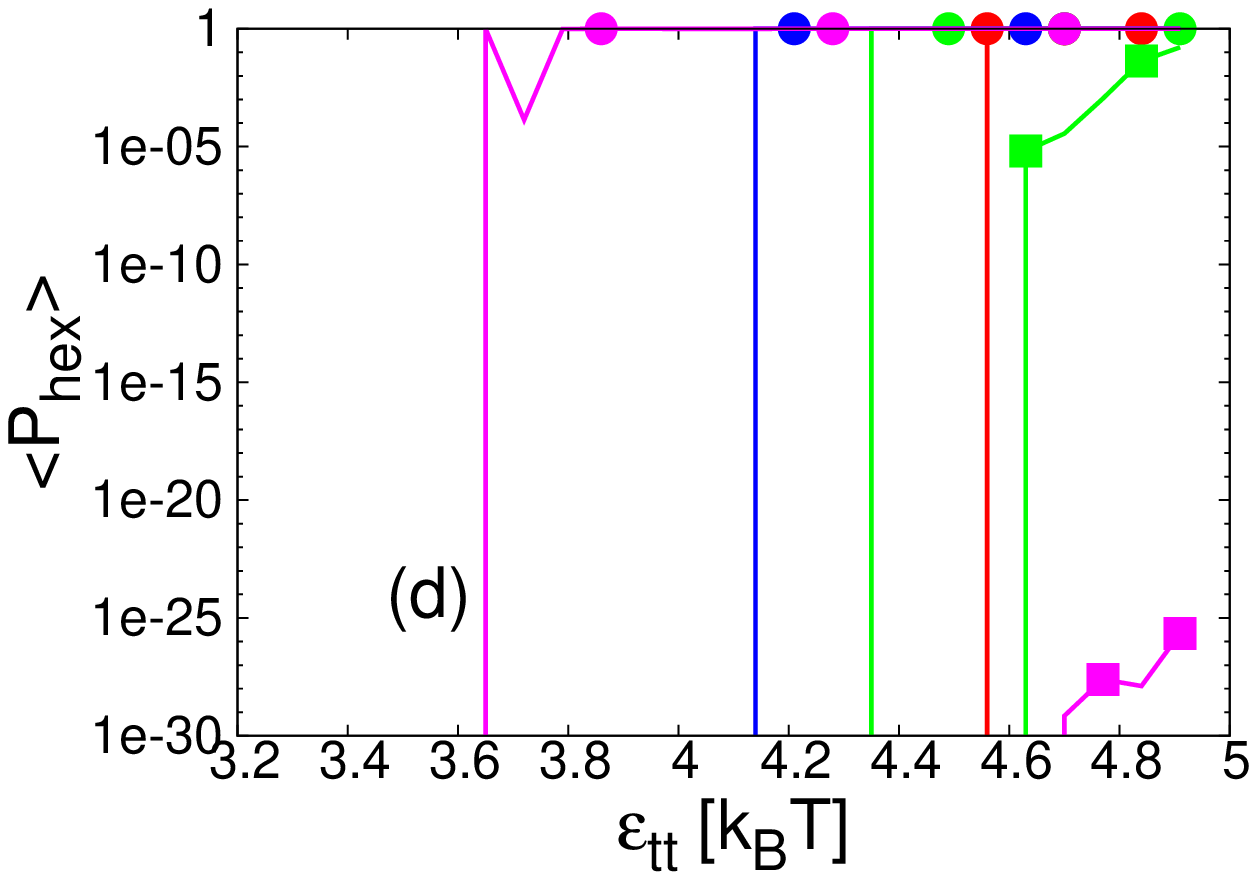}
\includegraphics[scale=0.4]{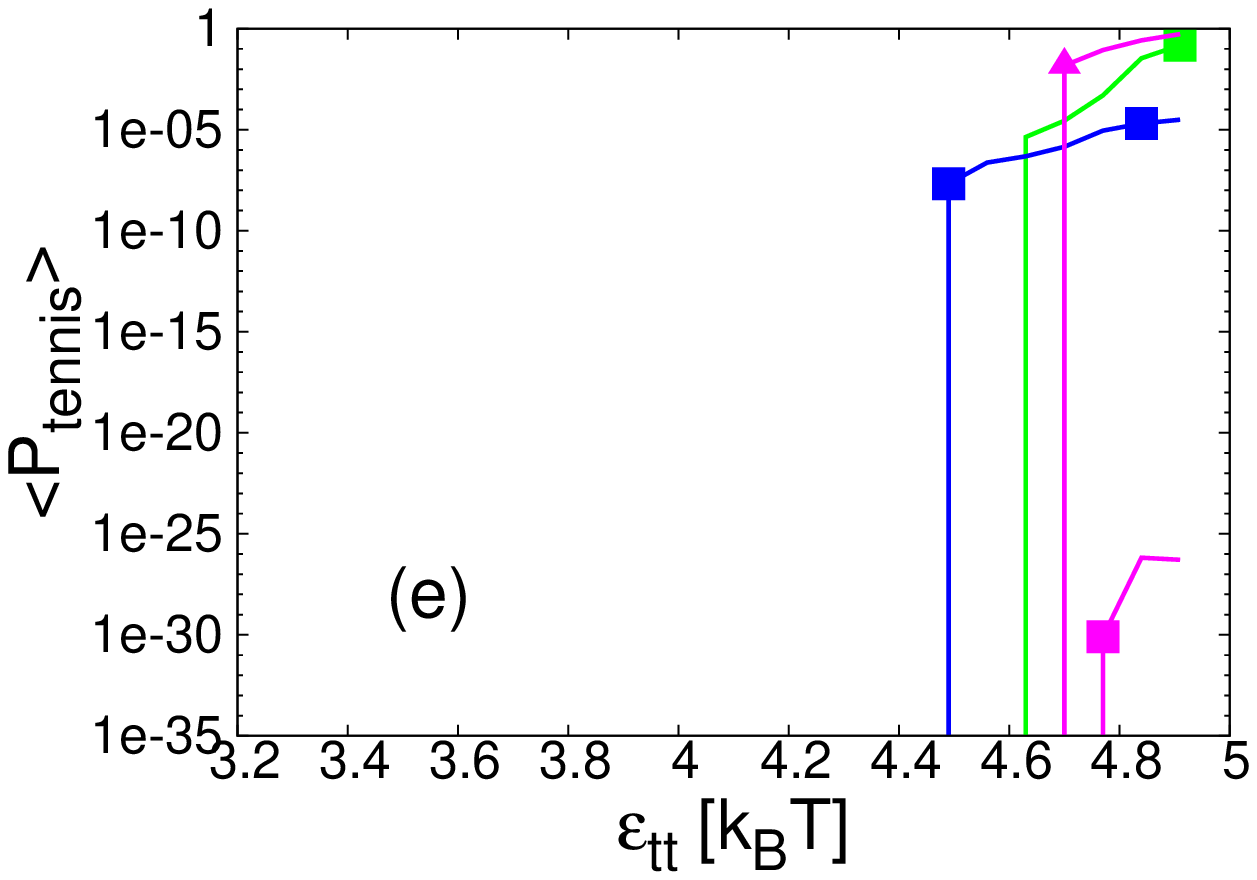}
\includegraphics[scale=0.4]{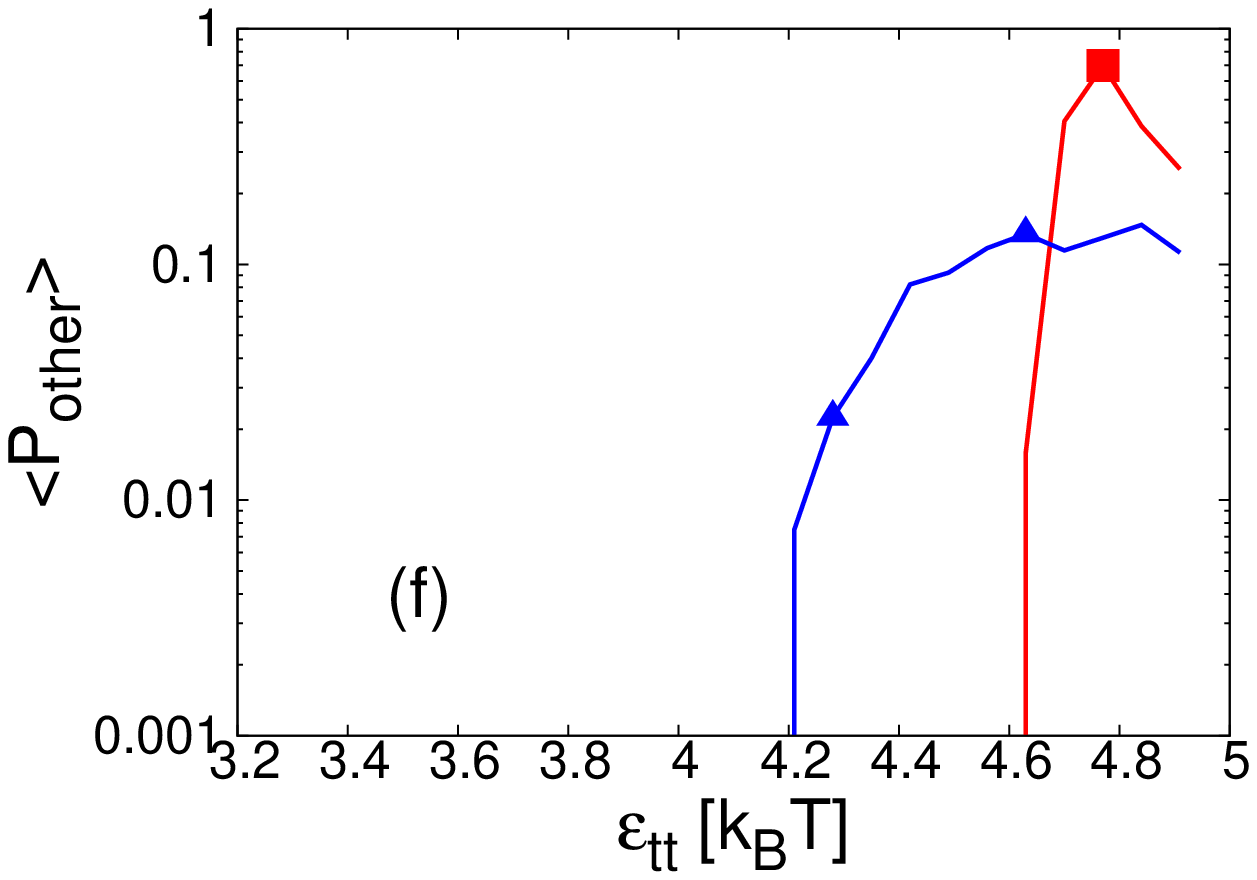}

\caption{\label{fig:MC_assembly} Average values as a function of $\epsilon_{tt}$ from single MC free assembly simulations with different flexibilities: $\kappa =  0.8 \times 10^3 k_BT$ (red), $\kappa =  1.6 \times 10^3 k_BT$ (green), $\kappa =  3.2 \times 10^3 k_BT$ (blue), rigid (magenta). Different simulation types: unassembled initial condition with umbrella sampling ($\blacksquare$); unassembled initial condition without umbrella sampling ($\blacktriangle$); assembled initial condition with umbrella sampling ($\bullet$). (a) Total inter-triskelion interaction energy. (b) Asphericity of the largest cluster in the system. (c) Mini-coat probability. (d) Hexagonal barrel probability. (e) Tennis ball probability. (f) Probability of different closed structure with the expected number of pentagons and hexagons, see text.}
\end{center}
\end{figure*}

We ran free assembly simulations with $N = 36$ triskelia, starting from two initial configurations: one with triskelia placed randomly, just with the requirement of no beads overlapping, and the other with an assembled hexagonal barrel. 24 systems with different $\epsilon_{tt}$ between $3.3 k_BT$ and $4.91 k_BT$ were run with parallel tempering swaps between them. In the majority of simulations, umbrella sampling with an iteratively-calculated weighting function~\cite{faller2002}, $w(U_{tt}/\epsilon_{tt})$, where $U_{tt}$ is the total inter-triskelion interaction energy, was used but some runs were also performed without. We first, in Fig.~\ref{fig:MC_assembly}, present results from individual free assembly simulations for various quantities as a function of $\epsilon_{tt}$.

\begin{figure}[ht!]
\begin{center}

\includegraphics[scale=0.4]{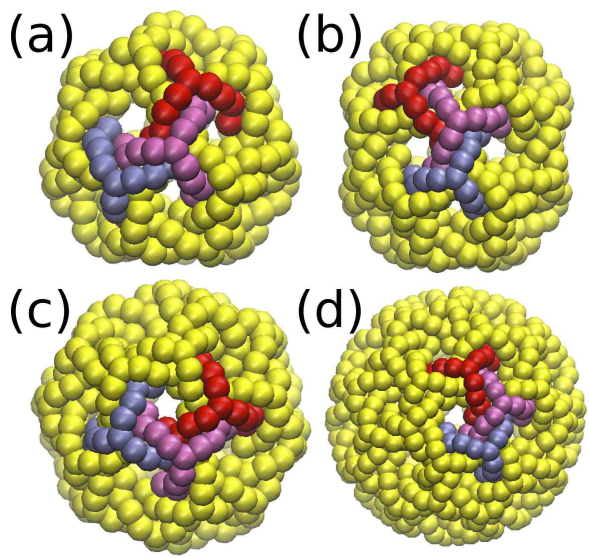}

\caption{\label{fig:cages_pics} Assembled structures: (a) Mini-coat with $N = 28$ triskelia (b) Hexagonal barrel, $N = 36$. (c) Tennis ball, $N = 36$. (d) Truncated icosahedron, $N = 60$. In each, three triskelia are shown in different colours to highlight their relative positions. The first three snapshots are all from free assembly simulations. The truncated icosahedron was not observed in free assembly so for the final configuration triskelia were placed by hand. The structure depicted is nonetheless mechanically stable for sufficient interaction strengths, see text and Fig.~\ref{fig:U_struct_eps_ss}.}
\end{center}
\end{figure}

Fig.~\ref{fig:MC_assembly}(a) shows the average inter-triskelion interaction energy, $\left<U_{tt}\right>$. For low $\epsilon_{tt}$, the results, at least for flexible triskelia, are similar for both assembled and unassembled initial configurations. However, for higher $\epsilon_{tt}$, the results for the assembled initial condition clearly show lower energies, indicating that proper sampling of equilibrium is not achieved. The point at which assembly starts is at higher $\epsilon_{tt}$ for more flexible triskelia, due to the larger loss of entropy. For the asphericity~\cite{aronovitz1986} of the largest cluster, $\left<\Delta \right>$, broadly similar results are seen for both initial conditions, with $\left<\Delta \right>$ changing from $\approx 1$ indicating highly aspherical structures when there is little assembly to  $\approx 0$ indicating almost spherical structures at high $\epsilon_{tt}$. In calculating the asphericity, the positions of the central, type $A$ beads were used. When there are no bonded triskelia in the system and thus the largest ``cluster'' is a single triskelion, the asphericity, which is calculated from a tensor based on the separations of pairs of beads~\cite{aronovitz1986}, is not defined. Since, when the largest cluster is of size 2, which is necessarily a line, $\Delta = 1$, we choose to assign a value of $\Delta = 1$ for a single trisklelion also.

Much larger differences are seen in Fig.~\ref{fig:MC_assembly}(c) - (f), where the probability of observing specific structures is considered. Here we plot the probabilities on logarithmic scales, down to very small values. These very low probabilities arise from the umbrella sampling: during the creation of the weighting function the system may become trapped in some configuration, eventually the weighting function will become large enough to allow the system to escape and explore other structures. However, due to the large weighting function, the estimated probability of these structures is very low. Since we know our simulations are not fully sampling equilibrium, these probabilities may well be severely underestimated. We check whether a bonded cluster has one of the four common structures - mini-coat, hexagonal barrel, tennis ball or truncated icosahedron - by using an algorithm~\cite{weinberg1966} to test if the graph formed by considering the bonds between triskelia is isomorphic to the one of the corresponding graphs. We identify two beads from different triskelia as being bonded if their interaction energy is $< -\frac{1}{4}\epsilon_{tt}$. Two triskelia are then defined to be bonded if there exists at least one bond between their type $B$ and type $C$ beads.

For the unassembled initial condition, we find that, overall, the most likely structure to be formed is the mini-coat, see Fig.~\ref{fig:MC_assembly}(c), both for simulations with and without umbrella sampling. At higher values of $\epsilon_{tt}$, some initially unassembled simulations did also form hexagonal barrel structures, consistent with the fact that interactions were chosen for this structure at low temperature, and also tennis ball structures, see Fig.~\ref{fig:MC_assembly}(d) and (e). In Fig.~\ref{fig:MC_assembly}(f) we show the probability of forming closed structures that have twelve pentagonal faces and $(N - 20)/2$ hexagonal faces, but which are not one of the known structures that we test for. These other structures that arose in our simulations had $32$ triskelia. No closed structures with a different number of pentagonal and hexagonal faces were formed, although there were additionally many open structures. For the assembled initial condition we found that the only closed structure seen in the simulation was the hexagonal barrel. We also ran free assembly simulations with 60 triskelia but no truncated icosahedra were assembled although, when pre-assembled, they were stable for higher $\epsilon_{tt}$. In Fig.~\ref{fig:cages_pics} we show snapshots of mini-coat, hexagonal barrel and tennis ball structures assembled in our simulations, as well as a truncated icosahedron structure.

\begin{figure}[ht!]
\begin{center}

\includegraphics[scale=0.6]{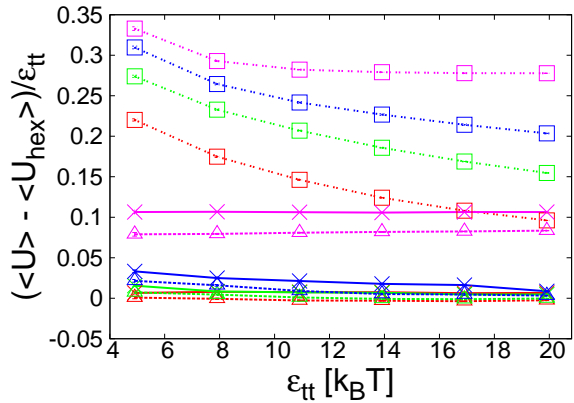}

\caption{\label{fig:U_struct_eps_ss} Difference between average internal energy per triskelion of a given structure - mini-coat ($\times$, solid line), tennis ball ($\triangle$, dashed line) or truncated icosahedron ($\square$, dot-dashed line)  -  and that of hexagonal barrel for different bending rigidities, $\kappa = 0.8 \times 10^3 k_BT$ (red), $\kappa = 1.6 \times 10^3 k_BT$ (green), $\kappa = 3.2 \times 10^3 k_BT$ (blue) and rigid (magenta). Note that for the lowest $\epsilon_{tt}$ for rigid triskelia, the mini-coat structure was unstable and disassembled: the data point plotted is only averaged over those parts of the simulations before disassembly occurred.}
\end{center}
\end{figure}

To obtain more information about the relative stability of the different structures, also for higher $\epsilon_{tt}$, we consider, in Fig.~\ref{fig:U_struct_eps_ss}, the average internal energy of a given structure divided by the number of triskelia in the structure, compared to the value for a hexagonal barrel. Simulations were run at single $\epsilon_{tt}$ values from $4.91 k_BT$ to $19.91 k_BT$ with only local moves. The initial condition was taken as the assembled structure and for all simulations, expect in one case, this structure persisted for the rest of the simulation: for the lowest $\epsilon_{tt}$ the structure was intermittently not identified according to our bonding definition, though only temporarily, indicating that true disassembly had not occurred. For rigid triskelia, the mini-coat with $\epsilon_{tt} = 4.91 k_BT$ was not stable. Although the overall structure did not disassemble, typically multiple bonds within the structure broke and did not reform within the simulation. It should be noted, however, that, given the mini-coat did persist for some time, and also given it was formed in some free assembly simulations with rigid triskelia, see Fig.~\ref{fig:MC_assembly}(c), at $\epsilon_{tt} = 4.91 k_BT$ the mini-coat must still represent a local minimum for the rigid triskelia.

We find that, usually, the hexagonal barrel has the lowest internal energy per triskelion, as expected since the parameters were determined for this structure. We find that the difference becomes more positive as the rigidity is increased but, for the highest two flexibilities, the tennis ball has lower internal energy per triskelion than the hexagonal barrel at some $\epsilon_{tt}$. The mini-coat has values relatively close to those for the hexagonal barrel but the truncated icosahedron has significantly higher values for all $\epsilon_{tt}$ considered. We observe that the difference relative to $\epsilon_{tt}$ decreases as $\epsilon_{tt}$ increases since the attractions dominate more over the bending rigidity but that for completely rigid triskelia the curves flatten out.

\begin{figure*}[ht!]
\begin{center}

\includegraphics[scale=0.9]{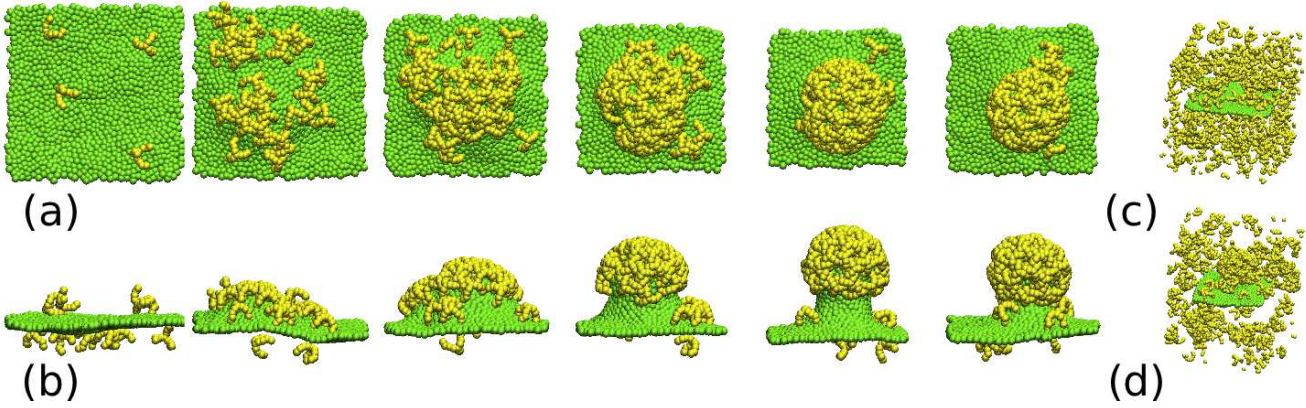}

\caption{\label{fig:assemb_on_mem} Snapshots for assembly on a membrane with $\kappa = 0.8 \times 10^3k_BT$, $\lambda_b = 8\sqrt{3} k_BT$, $\epsilon_{tt} = 5 k_BT$ and $\epsilon_{mt} = 2 k_BT$. Triskelia are depicted in yellow and membrane particles in green. For (a) and (b) only triskelia whose central bead is within $4\sigma$ of a membrane particle are shown: (a) Series of snapshots at regular time intervals from $t = 0 - 2.5\times10^4 t_0$. (b) Same configurations shown from different viewing angle. (c) Snapshot of same system at $t = 0$ with all triskelia. (d) Snapshot of same system at $t =  2.5\times10^4 t_0$ with all triskelia.}
\end{center}
\end{figure*}

\section{\label{sec:mem}Self-assembly on a membrane: hydrodynamics and bud formation}

Clathrin is intrinsically linked to membranes, and {\it in vivo} it is here that its self-assembly occurs~\cite{brodsky}. As discussed in Sec.~\ref{sec:intro}, the structures formed on a membrane are more poly-disperse than those assembled in the bulk and it is expected that the fluctuating surface will also change the pathways to assembly. In this section, we consider the assembly of our triskelia when they are attracted to a fluctuating membrane. In the terms used in the previous section, all the simulations with a membrane in this section are free assembly ones. We represent the membrane as a dynamically triangulated surface composed of bonded particles with a typical bond length of $\sigma$~\cite{noguchi2005}. Membrane fluidity is included by MC moves, performed at regular intervals, that attempt to flip bonds between neighbouring particles. The rate of bond-flipping sets the viscosity of the membrane, which may be measured by considering a Poiseuille flow in a two-dimensional membrane sheet, see appendix~\ref{app:mem}~\cite{noguchi2005}. 

Since free assembly simulations without a membrane indicated that our MC approach is unable to fully sample equilibrium, we proceed directly to dynamical simulations. We perform molecular dynamics simulations with our triskelia, as well as the membrane, coupled to a Stochastic Rotation Dynamics (SRD) solvent~\cite{gompper2009}. SRD is a coarse-grained method in which the fluid is represented by point particles of mass $m$ whose interactions are effected by dividing the system into a grid of cells at regular time intervals and exchanging momentum by a rotation through a certain angle of velocities relative to the cell centre of mass velocity. This acts as a thermostat, whilst also conserving momentum so that hydrodynamic interactions are included. More details of parameter choices and solute-solvent couplings are given in appendix~\ref{app:SRD}. Hydrodynamic interactions are naturally present in real, experimental systems and may have both qualitative and quantitative effects~\cite{kikuchi2002}. Although we do not investigate the effect of hydrodynamic interactions in detail, we choose a simulation method that includes them, as this is expected to be more dynamically realistic.

Since SRD requires the simulation box to be regularly divided into a grid with an integer number of cells, it is incompatible with the approach to simulating a tensionless membrane employed in our previous work without solvent~\cite{matthews2012} that involved box-rescaling. We therefore have developed a new approach, detailed in appendix~\ref{app:mem}, in which the edge of the membrane is bonded to a square frame, whose sides are a distance $r_{frame}$ from the edges of the simulation box. The frame may expand and contract. 

We define a unit of simulation time, $t_0 = \sigma \sqrt{k_BT/m}$. Our parameter choices give a membrane viscosity of $\eta_m = 35.1 \pm 0.1 m/ t_0$ and a fluid viscosity of $\eta_f = 2.5 m/\sigma t_0$, see appendices~\ref{app:mem} and~\ref{app:SRD}. The ratio of the viscosities is $l_{\eta} = \eta_m/\eta_f \approx 14 \sigma$. For a lipid bilayer in water, $l_{\eta} = 1 - 10 \mu$m~\cite{harland2010}. The typical size of a triskelion is on the order of $0.1 \mu$m, whereas in our simulation it is a few $\sigma$. Thus the size of a triskelion compared to $l_{\eta}$ is close to the lower end of the expected range, allowing efficient simulation.

As for the MC simulations, we consider triskelion stiffnesses of $\kappa = 0.8 \times 10^3  k_BT$, $1.6 \times 10^3  k_BT$ and $3.2 \times 10^3 k_BT$ and we consider membrane bending stiffnesses of $\lambda_b = 2\sqrt{3} k_BT$, $4\sqrt{3} k_BT$ and $8\sqrt{3} k_BT$. See appendix~\ref{app:mem} for the definition of the bending potential. Interactions of clathrin triskelia with membranes occur at the ends of the triskelion legs~\cite{kirchhausen2000} via intermediary adaptor proteins. We neglect the adaptor proteins and simply introduce an attractive interaction between the final beads in the legs and membrane particles, with a minimum of $-\epsilon_{mt}$. Unlike for triskelion-triskelion interactions, this attraction is not patchy but the bead-bead and membrane-bead potentials share a common radial form. We consider $\epsilon_{mt} = k_BT$ and $2 k_BT$, and $\epsilon_{tt} = 5 k_BT$ and $10 k_BT$.

We simulate 300 triskelia, $1156$ membrane particles and $\approx 5 \times 10^5$ SRD particles in a box of $45\sigma \times 45 \sigma \times 45 \sigma$ with periodic boundaries. The relatively high triskelion density, about $10$ times that used in previous work~\cite{otter2010}, is chosen such that assembly proceeds quickly but we do not expect it to qualitatively affect assembly on the membrane. An equilibration period with purely repulsive interactions of $3 \times 10^3 t_0$, chosen to be sufficient to allow the membrane to relax, was allowed before the system was simulated with attractions for $2.5 \times 10^{4} t_0$.

\begin{figure}[ht!]
\begin{center}

\includegraphics[scale=0.4]{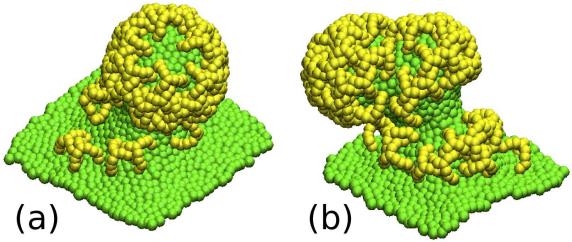}

\caption{\label{fig:defect_assembly} Snapshots of assembly on a membrane at $t = 2.5 \times 10^4 t_0$ with $\kappa = 0.8 \times 10^3k_BT$, $\epsilon_{tt} = 5 k_BT$ and $\epsilon_{tt} = 2 k_BT$. Colouring as in Fig.~\ref{fig:assemb_on_mem}, only triskelia whose central bead is within $4\sigma$ of a membrane particle are shown. (a) $\lambda_b = 8\sqrt{3} k_BT$, showing an example of assembly with a gap in the cage. (b) $\lambda_b = 4\sqrt{3} k_BT$, showing an example of assembly with two distinct cages forming a double-headed structure.}
\end{center}
\end{figure}

For most parameter choices, the triskelia assembled on the the membrane, causing the membrane to form a bud, see Fig.~\ref{fig:assemb_on_mem}. The example in Fig.~\ref{fig:assemb_on_mem} shows the formation of a relatively defect-free, approximately spherical cage on the membrane. Often, however, the cages formed had defects or gaps of varying sizes, see Fig.~\ref{fig:defect_assembly}(a). Additionally, some runs produced structures with two largely separate cages attached to one bud, causing a double-headed structure, see Fig.~\ref{fig:defect_assembly}(b). Similarly lumpy structures have been observed experimentally in clathrin assembly~\cite{iversen2003}.

\begin{figure}[ht!]
\begin{center}

\includegraphics[scale=0.6]{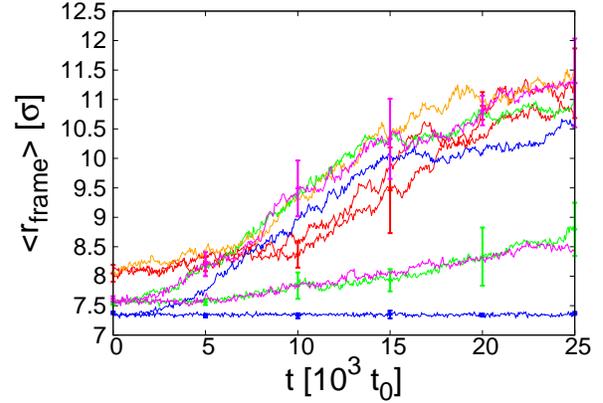}

\caption{\label{fig:r_frame_time} Average over 5 independent runs of $r_{frame}$ against time for different parameters. Errorbars show the standard deviation of the data. They are only plotted intermittently and for some curves for clarity. Red: $\kappa = 0.8 \times 10^3 k_BT$, $\lambda_b =  2\sqrt{3} k_BT$ for $\epsilon_{mt} = k_BT$, $\epsilon_{tt} = 5 k_BT$ (with errorbars) and $\epsilon_{mt} = k_BT$, $\epsilon_{tt} = 10 k_BT$ (without errorbars). Orange: $\kappa = 0.8 \times 10^3 k_BT$, $\lambda_b =  2\sqrt{3} k_BT$, $\epsilon_{mt} = 2 k_BT$, $\epsilon_{tt} = 5 k_BT$. Green: $\kappa = 0.8 \times 10^3 k_BT$, $\lambda_b =  4\sqrt{3} k_BT$ for  $\epsilon_{mt} = k_BT$, $\epsilon_{tt} = 5 k_BT$ (with errorbars) and $\epsilon_{mt} = 2 k_BT$, $\epsilon_{tt} = 10 k_BT$ (without errorbars). Blue: $\kappa = 0.8 \times 10^3 k_BT$, $\lambda_b =  8\sqrt{3} k_BT$ for  $\epsilon_{mt} = k_BT$, $\epsilon_{tt} = 10 k_BT$ (with errorbars) and $\epsilon_{mt} = 2 k_BT$, $\epsilon_{tt} = 10 k_BT$ (without errorbars). Magenta: $\kappa = 3.2 \times 10^3 k_BT$, $\lambda_b =  4\sqrt{3} k_BT$ for $\epsilon_{mt} = 2 k_BT$, $\epsilon_{tt} = 10 k_BT$ (with errorbars) and $\epsilon_{mt} = k_BT$, $\epsilon_{tt} = 5 k_BT$ (without errorbars).}
\end{center}
\end{figure}

We found that $r_{frame}$ was a good indicator of whether budding had occurred, moving to higher values as membrane area was taken into the bud and the frame contracted. In Fig.~\ref{fig:r_frame_time} we plot $\left<r_{frame}\right>$ against time for a variety of parameters. We first note that the results for different $\kappa$ are similar and also that the rate of bud formation did not show strong dependence on $\epsilon_{tt}$. In contrast, the rate of bud formation did depend on the values of $\epsilon_{mt}$ and $\lambda_b$. 

For the stronger attraction of the triskelia to the membrane, $\epsilon_{mt} = 2 k_BT$, the rate of bud formation was similar all $\lambda_b$ considered and the results were also similar for $\epsilon_{mt} = k_BT$ with $\lambda_b = 2\sqrt{3} k_BT$. However, with the weaker attraction to the membrane, $\epsilon_{mt} = k_BT$, when the stiffness of the membrane was increased to $\lambda_b = 4\sqrt{3} k_BT$, the rate was significantly slower, although clear buds were formed. Increasing the membrane stiffness further to $\lambda_b = 8\sqrt{3} k_BT$, again with $\epsilon_{mt} = k_BT$, no clear buds were formed within $2.5 \times 10^4 t_0$, although caps on the membrane with some curvature were formed in some runs. The assembly of extended flat sheets as in previous work~\cite{matthews2012} did not occur. 

\begin{figure}[ht!]
\begin{center}

\includegraphics[scale=0.6]{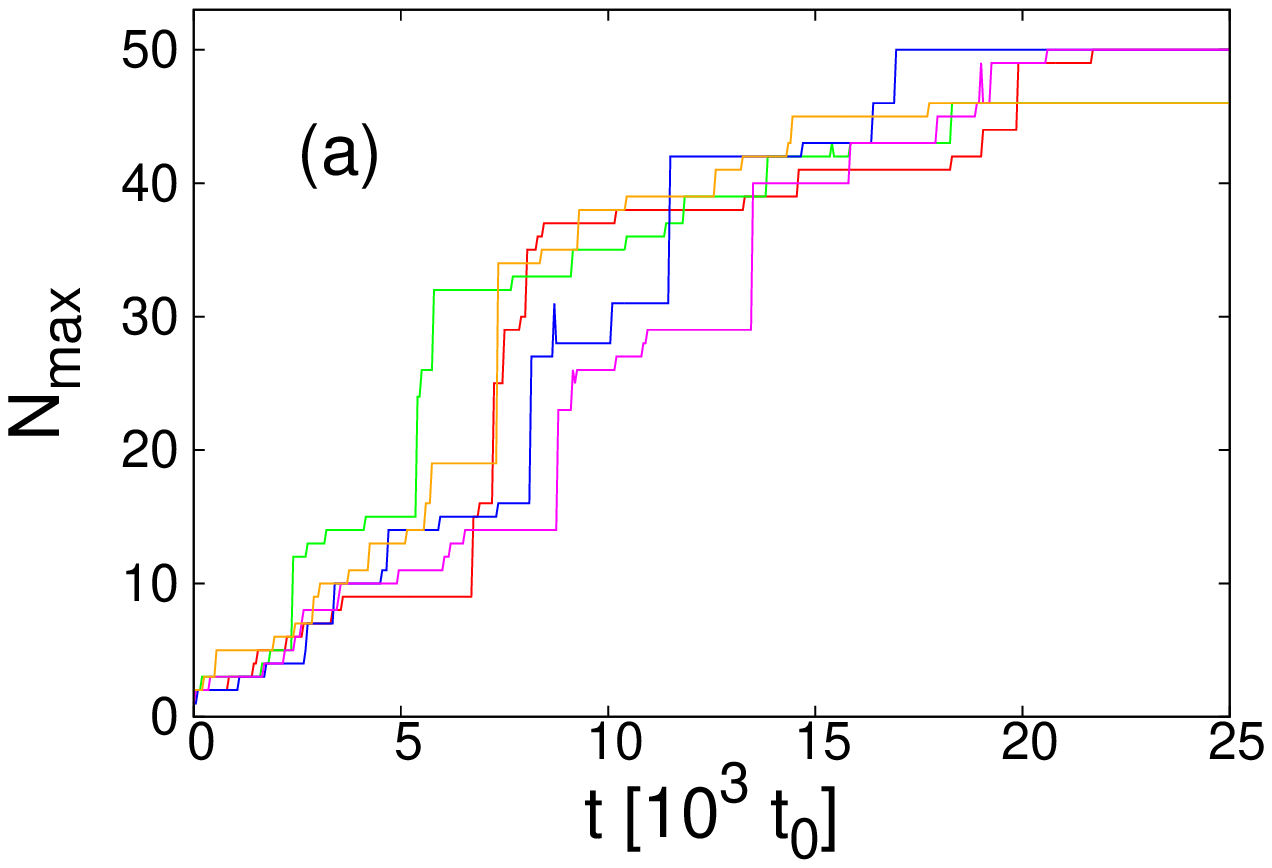}
\includegraphics[scale=0.6]{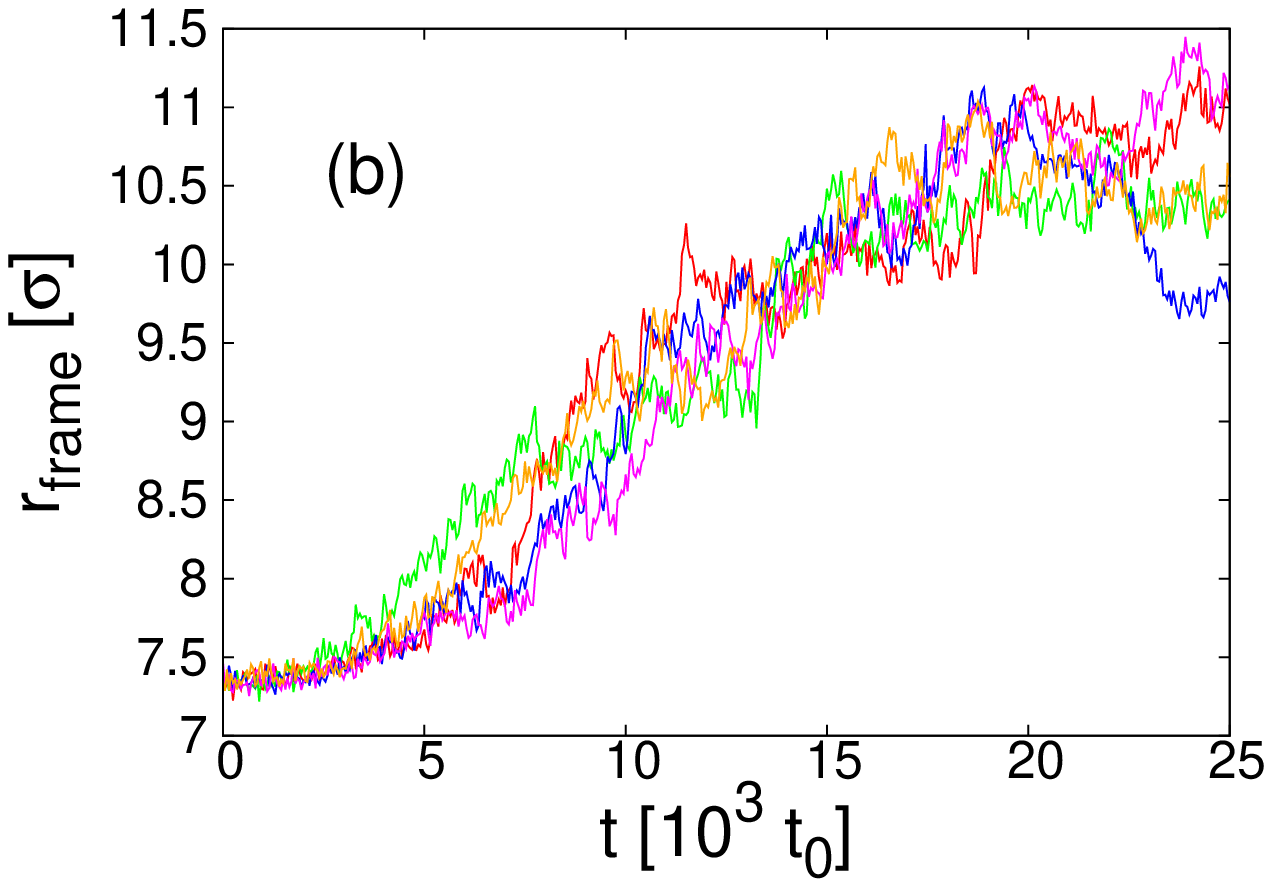}

\caption{\label{fig:r_frame_indiv} Individual runs for $\kappa = 0.8 \times 10^3 k_BT$, $\lambda_b =  8\sqrt{3} k_BT$ for  $\epsilon_{mt} = 2 k_BT$ , $\epsilon_{tt} = 5 k_BT$. (a) The population of the largest cluster in the system, $N_{max}$, against time. (b) $r_{frame}$ against time. Curves of the same colour in the different plots show data for the same run. The blues curves show data for the run that is depicted in snapshots in Fig.~\ref{fig:assemb_on_mem}.}
\end{center}
\end{figure}

A typical pathway to bud formation was for multiple smaller clusters to form on the membrane, see for example the second snapshots in Fig.~\ref{fig:assemb_on_mem}(a) and (b), and then coalesce, leading to a more rapid increase in the membrane curvature and bud formation, see for example the third and fourth snapshots in Fig.~\ref{fig:assemb_on_mem}(a) and (b). For many runs, though not all, the footprint of this pathway could be seen by comparing the number of triskelia in the largest cluster in the system, $N_{max}$, and $r_{frame}$ as a function of time. As may be seen by comparing Fig.~\ref{fig:r_frame_indiv}(a) and Fig.~\ref{fig:r_frame_indiv}(b) the most rapid increase in $r_{frame}$ is correlated with a rapid increase in $N_{max}$, corresponding to smaller clusters joining together.

\begin{figure}[ht!]
\begin{center}

\includegraphics[scale=0.6]{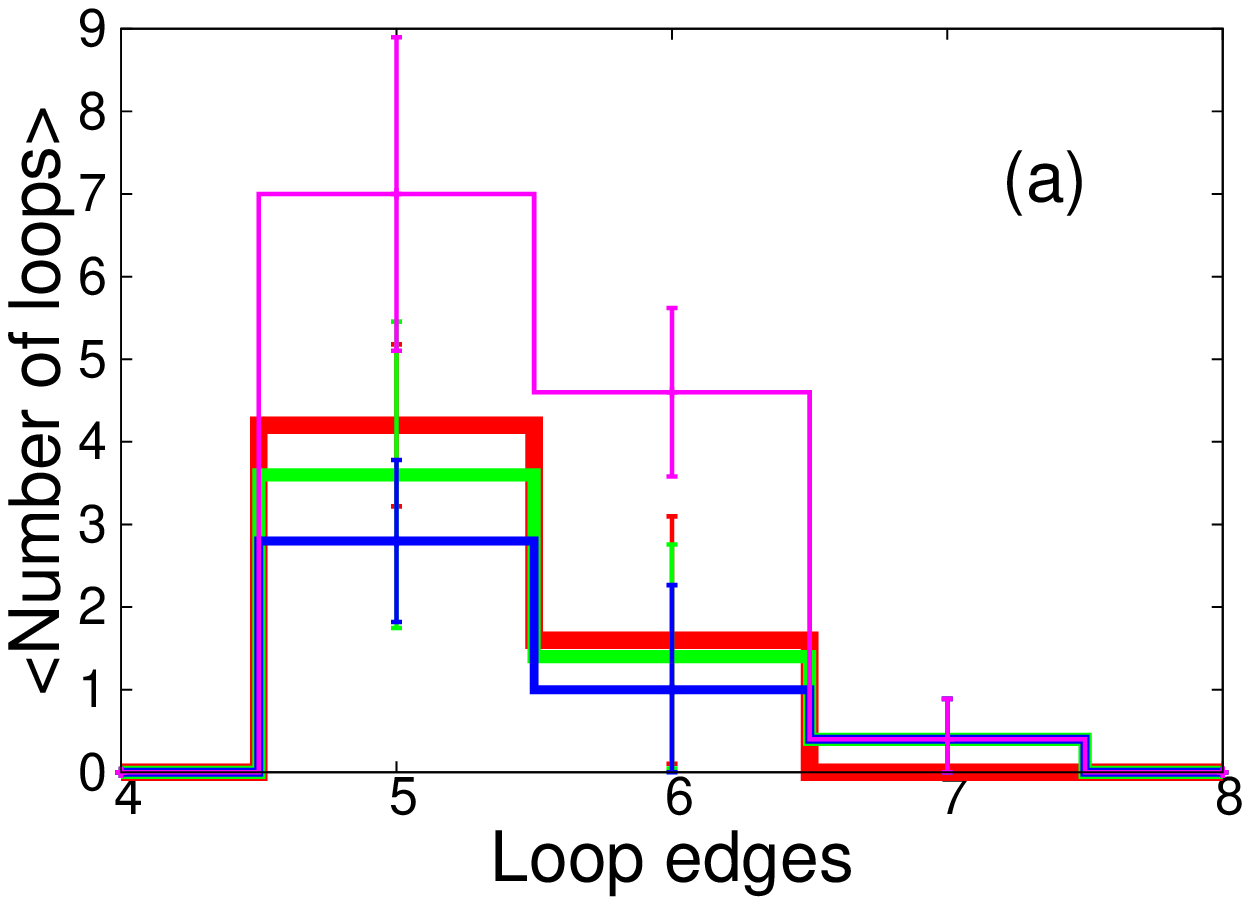}
\includegraphics[scale=0.6]{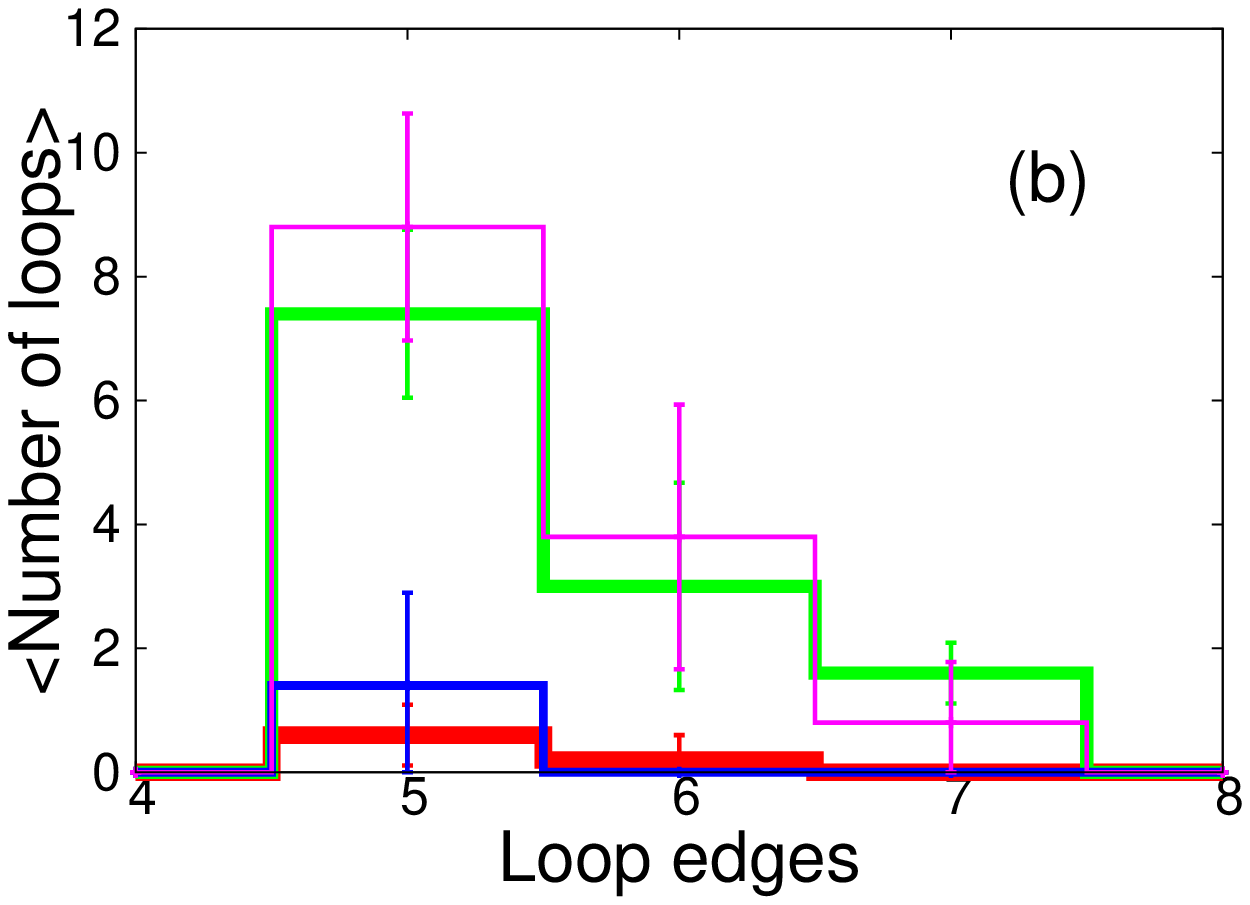}

\caption{\label{fig:Loop_hist_2_300_800} Histograms of the number of closed loops surrounding faces formed by the triskelia with different numbers of edges at $t = 2.5 \times 10^4 t_0$. Averaged over 5 independent runs, errorbars show standard deviations. (a) $\kappa = 0.8 \times 10^3 k_BT$, $\lambda_b =  2\sqrt{3} k_BT$ (b)  $\kappa = 0.8 \times 10^3 k_BT$, $\lambda_b =  8\sqrt{3} k_BT$. For both plots the different colours denote the same $\epsilon_{tt}$ and $\epsilon_{mt}$ values: $\epsilon_{tt} = 5 k_BT$, $\epsilon_{mt} = k_BT$ (red); $\epsilon_{tt} = 5 k_BT$, $\epsilon_{mt} = 2 k_BT$ (green); $\epsilon_{tt} = 10 k_BT$, $\epsilon_{mt} = k_BT$ (blue); $\epsilon_{tt} = 10 k_BT$, $\epsilon_{mt} = 2 k_BT$ (magenta). The thickness of the lines is varied for clarity.}
\end{center}
\end{figure}

We finally, in Fig.~\ref{fig:Loop_hist_2_300_800}, plot the distribution of the number of edges in the closed loops surrounding faces formed by assembled triskelia at the end of the simulation. It should be noted that this includes a contribution from assembly in the bulk as well as on the membrane, although this should be similar for all $\epsilon_{mt}$ and $\lambda_b$ and was small, as may be seen from the results for $\epsilon_{mt} = k_BT$ in Fig.~\ref{fig:Loop_hist_2_300_800}(b). Although large variation was seen, generally more loops with 5 edges than loops with 6 edges were formed. For some parameters, 7-edge loops were formed.

\section{\label{sec:conc}Conclusions}

We have introduced a new clathrin model using patchy beads that allows the inclusion of excluded volume and flexibility, as well as the interweaving of triskelia in assembled structures. Further, we have also described an approach to producing parameters for the model that will allow the assembly of similar structures to those seen in nature. Choosing the hexagonal barrel as a target structure, we employed our approach to find a parameter set. MC simulations using these parameters showed that the triskelia could assemble a hexagonal barrel, as well as other structures observed in nature: the mini-coat and the tennis ball. Additionally, further structures were formed with different numbers of triskelia but also with 12 pentagonal and $(N - 20)/2$ hexagonal faces, where $N$ is the number of triskelia. The MC simulations were found to not be able to access all of the various local minima within one run. The mini-coat, hexagonal barrel,  tennis ball, as well as truncated icosahedron structures were found to be mechanically stable for a range of triskelion stiffnesses, for sufficient attraction strengths. 

Dynamical simulations of the assembly of the model triskelia with an attractive fluctuating membrane, employing a new membrane boundary condition, were performed with coupling to a coarse-grained solvent to include hydrodynamic interactions. For most parameters, the formation of buds by the assembly of the triskelia on the membrane surface was found. The buds were surrounded by cages with pentagonal, hexagonal and sometimes heptagonal faces. They often contained defects or holes and sometimes had lumpy, double-headed structures.

Our model takes into consideration key characteristics of clathrin, such as excluded volume, flexiblity and binding site selectivity, whilst at the same time remaining computationally tractable. It is capable of reproducing the salient observed features of the protein: the assembly and stability of known structures and the formation of buds on a membrane. Whilst the smaller number of beads used in the current model is advantageous for simulation, its success suggests it could be interesting in further work to consider a similar model with a finer coarse-graining, that might be able to capture even more features. A similar approach might also be applied to some of the other proteins that attach to the membrane during budding~\cite{liu2009}, and it could be very interesting to model their collaborative binding.

This work was supported by the Austrian Science Fund (FWF): M1367. Snapshots were created using VMD~\cite{humphrey}. The computational results presented have been achieved in part using the Vienna Scientific Cluster (VSC).

\appendix

\section{\label{sec:app}Appendix}

We present additional details of our model and methods. Many features of our model are similar to our previous work~\cite{matthews2012} and, correspondingly, parts of the descriptions in this appendix are very similar to parts of the supplemental material in ref.~\cite{matthews2012}. They are nonetheless reproduced here for the convenience of the reader.

\subsection{\label{app:trisk-trisk} Triskelion-triskelion interactions}
We first discuss the form of the triskelion-triskelion, $tt$, interactions, which have the same radial form as the triskelion-membrane particle, $mt$, interactions. The potential form is similar to that used in earlier work~\cite{wilber}. For two different particles, $i$ and $j$, separated by $r_{ij} = |\mathbf{r}_{ij} | =  | \mathbf{r}_j - \mathbf{r}_i | $, where $\mathbf{r}_{i}$  is position of particle $i$, the general form for both these types of interactions is, 
\begin{eqnarray}
\nonumber U_{ij}&=& \gamma_{area} \left[U_{WCA}(r_{ij}) + \gamma_{att} \gamma_{orient} U_{att}(r_{ij}) \right]
\\\nonumber U_{WCA}(r)&=& 
\left\{
\begin{array}{l}
4\epsilon\left[\left(\frac{\sigma}{r}\right)^{12}-\left(\frac{\sigma}{r}\right)^{6} + \frac{1}{4} \right]\\
\hspace{1.2 in} \mathrm{for} \; r < r_t,  \\
0\\
\hspace{1.2 in} \mathrm{for} \; r \ge r_t, \\
\end{array}
\right.
\\U_{att}(r)&=& 
\left\{
\begin{array}{l}
-\epsilon\\
\hspace{1.2 in} \mathrm{for} \; r < r_t,\\
4\epsilon\left[\left(\frac{\sigma}{r}\right)^{12}-\left(\frac{\sigma}{r}\right)^{6} \right]\\
\\
\hspace{1.2 in} \mathrm{for} \; r_t  \le r \le r_s,\\
a(r - r_c)^2 + b(r - r_c)^3\\ 
\\
\hspace{1.2 in} \mathrm{for} \; r_s \le r \le r_c,\\
0 \\
\hspace{1.2 in} \mathrm{for} \; r \ge r_c,\\
\end{array}
\right.
\label{eq:LJ_pot}
\end{eqnarray}
where $r_t = 2^{1/6}\sigma$, $r_s = (\frac{26}{7})^{1/6}\sigma$, $r_c = \frac{67}{48}r_s$, $a = - \frac{24192}{3211}\frac{\epsilon} {r_s^2}$ and $b = -\frac{387072}{61009}\frac{\epsilon}{r_s^3}$. The form of $U_{att}(r)$ in the range $ r_s \le r \le r_c$ is a polynomial interpolation used to avoid a jump in the potential or its first derivative at the cut-off~\cite{bordat2001}. The energy scale, $\epsilon$, is set to $\epsilon_{tt}$ for triskelion-triskelion interactions. The dimensionless factors $\gamma_{area}$, $\gamma_{att}$ and $\gamma_{orient}$ take different forms for $tt$ and $mt$ interactions.

For $tt$ interactions $\gamma_{area} = 1$. The patches on triskelion beads are given identities and only specific pairs are attractive, as detailed in the main text. For a pair of interacting triskelion beads from different triskelia, if the pair of closest patches are attractive then $\gamma_{att} = 1$, otherwise $\gamma_{att} = 0$.

\begin{figure}[ht!]
\begin{center}

\includegraphics[scale=0.8]{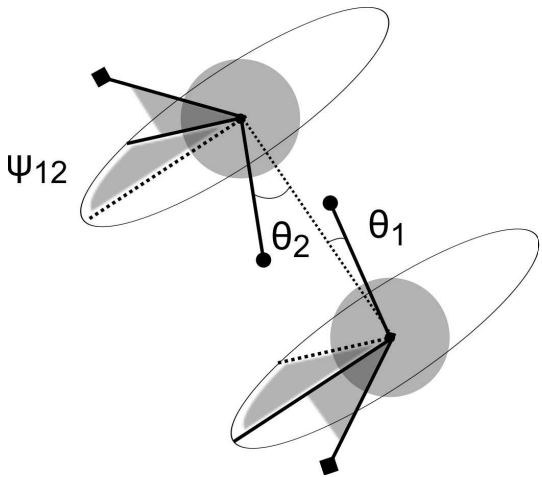}

\caption{\label{fig:bead_internal_interact} Sketch to illustrate the use of the unit vectors associated with bonded beads in a triskelion in calculating the internal interactions between those beads. The bending potential depends on the angles between the vectors terminating in circles, denoted bending vectors, and the bead-to-bead vector. The torsion potential depends on the angles between the projections of the two vectors terminating in diamonds, denoted internal torsional vectors, onto the plane perpendicular to the bead-to-bead vector. For a given bead of type $\alpha$ the relative directions of the different vectors for internal interactions are defined by the $\left\{\mathbf{u}[\alpha,i]\right\}$ parameters.}

\end{center}
\end{figure}

The factor $\gamma_{orient}$ allows the attractive part of the interaction to be made patchy. For $tt$-interactions the centres of the attractive patches are defined by unit vectors. For a given bead of type $\alpha$, the relative directions of these are given by the $\left\{\mathbf{v}[\alpha,i]\right\}$ parameters. The width of the patches is determined by $\gamma_{orient}$, which is a product of functions of the form~\cite{miller2009}:
\begin{equation}
F(\phi; \phi_a, \phi_b) = 
\left\{
\begin{array}{l}
1\\
\hspace{1.2 in} \mathrm{for} \; \phi \le \phi_a, \\
\cos^2[(\pi / 2) (\phi- \phi_a) / \phi_b]\\
\\
\hspace{1.2 in} \mathrm{for} \; \phi_a \le \phi \le \phi_a  + \phi_b, \\
0 \\
\hspace{1.2 in} \mathrm{for} \; \phi \ge  \phi_a  + \phi_b.\\
\end{array}
\right.
\end{equation} 
$\gamma_{orient}(\hat{\mathbf{r}}_{ij}, \mathbf{\Omega}_i,\mathbf{\Omega}_j) =  F(\theta_i;\theta_a,\theta_b) \times F(\theta_j;\theta_a,\theta_b) \times F(\psi_{ij}; 2\theta_a,2\theta_b)$, where $\mathbf{\Omega}$ describes particle orientation.  $\theta_i$ is the angle between the interacting patch on particle $i$ and $\hat{\mathbf{r}}_{ij}$, whilst $\theta_j$ is between the patch of particle $j$ and $-\hat{\mathbf{r}}_{ij}$. $\psi_{ij}$ is the angle between the projections of the external torsional vectors of $i$ and $j$ onto the plane perpendicular to $\hat{\mathbf{r}}_{ij}$. The factor $F(\psi_{ij})$ penalizes the twisting of interacting sub-units. We follow ref.~\cite{wilber} in choosing the range for this factor to be double that for the other ones. We choose $\theta_a = 0.3$ and $\theta_b = 0.2$. The same geometry as for the external patchy interactions is also used for the internal interactions, see Fig.~\ref{fig:bead_internal_interact}.

\subsection{\label{app:trisk_intern} Triskelion internal interactions}

We next define the internal interactions between bonded beads within the same triskelion. Similarly to the external attractive patchy interactions, these are based on unit vectors associated with the bonded bead types. For a given bead of type $\alpha$ their relative directions are given by the $\left\{\mathbf{u}[\alpha,i]\right\}$ parameters. Fig.~\ref{fig:bead_internal_interact} depicts how the vectors are used to determine a set of angles between a pair of bonded beads. For these internal interactions, we denote the vectors whose angles to the bead-to-bead vector are considered as bending vectors, whilst the vectors whose rotations around it are considered are denoted internal torsional vectors. It should be noted that the torsional vectors for external and internal interactions are not the same. 

A potential is then applied, $U_{internal} = \frac{1}{2}k (r - d)^2 + \kappa(1 - cos(\theta_1)) + \kappa(1 - cos(\theta_2)) + \kappa(1 - cos(\psi_{12}))$, where $r$ is the bead separation, $\theta_1$ and $\theta_2$ are angles between the bending vectors from the two beads respectively and the centre-to-centre vector. $\psi_{12}$ is the angle between the projections of the two internal torsional vectors from the beads onto the plane perpendicular to the bead-to-bead vector.

\subsection{\label{app:param_det} Triskelion parameter determination}

\begin{figure}[ht!]
\begin{center}

\includegraphics[scale=0.43]{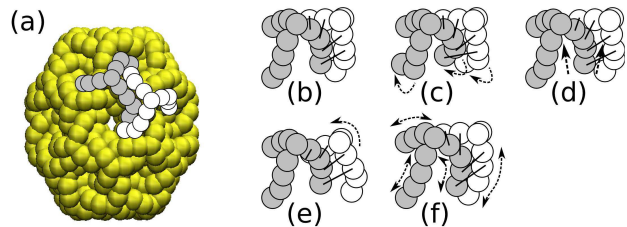}

\caption{\label{fig:clath_find_interact_with_hex} Sketch of the process used to find a triskelion shape and set of interactions. (a) 36 triskelia are arranged with their centres on the vertices of a hexagonal barrel. Two neighbouring triskelia are highlighted in white and grey: (b) The pairs of beads that interact are indicated by thick connecting lines. Low-temperature Monte Carlo simulations are performed with a range moves: (c) A vector, $\mathbf{u}^{*}$, that determines the internal triskelion interactions is changed. Note that this changes the shape of all legs on all triskelia. Vectors defining the patches for external interactions, $\mathbf{v}^{*}$ are also changed (not depicted). (d) The edge length, $l_{hex}$ of the hexagonal barrel is varied, all triskelia receive a corresponding radial displacement. (e) Individual triskelia are rotated around their centre. (f) The separation of beads along the legs, $d$, is changed. In the latter stages of the simulation, triskelion centres are allowed to move freely.}
\end{center}
\end{figure}

\begin{table*} [ht!]
\begin{center}
\begin{tabular}{| p{0.7cm}  | p{0.7cm}  | p{0.7cm}  | p{0.7cm}  | p{1.4cm} | p{1.4cm} | p{1.4cm} | } \hline

\multicolumn{2}{|c|}{Vector identity} & \multicolumn{2}{c|}{Bonding to} & \multicolumn{3}{c|}{Vector components} \\ \hline

$\alpha$ & $i$ & $\beta$ & $j$ & $x$ & $y$ & $z$ \\ \hline

\multirow{4}{*}{$A$}&1 & $B$ &1 &0.95812 &-0.28261 &0.04633\\ \cline{2 - 7}
&2 & $B$ &1&-0.68000 &-0.68086 &-0.27208   \\  \cline{2 - 7}
&3 & $B$ &1&-0.19950 &0.96347 &-0.17868   \\  \cline{2 - 7}
&4 & \multicolumn{2}{c|}{{\it torsion}} &0.19081 &0.00000 & -0.98163\\ \hline

\multirow{3}{*}{$B$}&1 & $A$ &1,2,3 &-0.95812 &0.28261 &-0.04633\\ \cline{2 - 7}
&2 & $C$ &1&0.80148 &-0.32975 &-0.49889   \\ \cline{2 - 7}
&3 & \multicolumn{2}{c|}{{\it torsion}} &-0.20924 &0.11629 & -0.97092\\ \hline

\multirow{3}{*}{$C$}&1 & $B$ &2 &-0.80148 & 0.32975 & 0.49889\\ \cline{2 - 7}
&2 & $D$ &1&0.73354 &0.08292 &-0.67456   \\ \cline{2 - 7}
&3 & \multicolumn{2}{c|}{{\it torsion}} &-0.55936 &0.26125 &-0.78668 \\ \hline

\multirow{3}{*}{$D$} &1 & $C$ &2 &-0.73354 &-0.08292 & 0.67456\\ \cline{2 - 7}
&2 & $E$ &1&0.14136 & -0.48848 & -0.86105   \\  \cline{2 - 7}
&3 & \multicolumn{2}{c|}{{\it torsion}} & -0.81931 & 0.22663 & -0.52665\\ \hline 

\multirow{2}{*}{$E$} &1 & $D$ &2 &-0.14136 &0.48848 &0.86105\\  \cline{2 - 7}
&2 & \multicolumn{2}{c|}{{\it torsion}} & -0.87421 &0.34653 & -0.34012 \\  \hline

\end{tabular}

\caption{\label{tab:u_vecs} The vectors, $\left\{\mathbf{u}[\alpha,i]\right\}$,  extracted from the parameter-finding simulations and employed in the internal interactions between bonded triskelion beads that determine the mechanical equilibrium shape in the free assembly simulations. The first two columns show the bead type, $\alpha$, and vector index, $i$. The second two columns show the bead type, $\beta$, and index, $j$ corresponding to the other bending vector involved in the interaction, or alternatively show the label {\it torsion}, indicating the vector is used in the torsional component of interactions. Each bead type only has one torsional vector for internal interactions, which is used for all such interactions. Vector components are given to 5 decimal places.}
\end{center}
\end{table*}

\begin{table*}  [ht!]
\begin{center}
\begin{tabular}{| p{0.7cm}  | p{0.7cm}  | p{0.7cm}  | p{0.7cm}  | p{1.4cm} | p{1.4cm} | p{1.4cm} | } \hline

\multicolumn{2}{|c|}{Patch identity} & \multicolumn{2}{c|}{Attracts} & \multicolumn{3}{c|}{Vector components} \\ \hline

$\alpha$ & $i$ & $\beta$ & $j$ & $x$ & $y$ & $z$ \\ \hline

\multirow{3}{*}{$B$}&1 & $C$ &1&0.49296 & 0.78900 &-0.36669 \\ \cline{2 - 7}
&2 & $D$ & 1 & -0.31730&-0.43760 &-0.84132  \\  \cline{2 - 7}
&3 & \multicolumn{2}{c|}{{\it torsion}}&0.84037 &-0.54084 &-0.03577\\  \hline

\multirow{3}{*}{$C$}&1 & $B$ &1 &0.52052 &0.79423 &0.31345\\ \cline{2 - 7}
&2 & $E$ &1&-0.40401 &-0.09107 &-0.91021  \\  \cline{2 - 7}
& 3 & \multicolumn{2}{c|}{{\it torsion}} &-0.84655 &0.53135 &0.03203\\  \hline

\multirow{3}{*}{$D$}&1 & $B$ &2 &0.82034 &-0.13387 & 0.55599\\  \cline{2 - 7}
& 2 & \multicolumn{2}{c|}{{\it torsion}} &-0.14006 &-0.97990 &-0.14203\\  \hline 

\multirow{2}{*}{$E$} &1 & $C$ &2 &0.62525 &-0.39778 &0.67144\\  \cline{2 - 7}
& 2 & \multicolumn{2}{c|}{{\it torsion}} &0.24275 &0.92872 &0.28026 \\  \hline

\end{tabular}
\caption{\label{tab:v_vecs} The vectors, $\left\{\mathbf{v}[\alpha,i]\right\}$, extracted from the parameter-finding simulations and employed in the external interactions between triskelion beads from different triskelia in the free assembly simulations. The first two columns show the bead type, $\alpha$, and patch index, $i$. The second two columns show the bead type, $\beta$, and index, $j$ corresponding to the other patch that the first patch attracts, or alternatively show the label {\it torsion}, indicating the vector is used in the torsional component of interactions. Each bead type only has one torsional vector for external interactions, which is used for all such interactions. Vector components are given to 5 decimal places.}
\end{center}
\end{table*}

We next give more details of the simulations used to determine triskelion parameters. 36 triskelia were placed with their centres on the vertices of a hexagonal barrel of edge length $l_{hex}$, although it should be noted that the hexagonal barrel structure cannot be formed from regular hexagons and pentagons~\cite{schein2008} and so there is some ambiguity as to the exact positions. Initial values for $\left\{\mathbf{v}^{*}[\alpha,i]\right\}$, $\left\{\mathbf{u}^{*}[\alpha,i]\right\}$ and $d$ were chosen by hand so that triskelia pairs interact in approximately the desired way. For the initial part of the simulation, triskelion centre positions remained fixed, except in updates of $l_{hex}$ with a corresponding radial displacement of all centres. Further moves, summarised in Fig.~\ref{fig:clath_find_interact_with_hex}, included varying $d$ between $0.5\times2^{1/6} \sigma$ and $1.3\times2^{1/6}\sigma$, as well as updates of $\left\{\mathbf{v}^{*}[\alpha,i]\right\}$ and $\left\{\mathbf{u}^{*}[\alpha,i]\right\}$, and rigid-body rotations around triskelion centres. Due to the uncertainty about vertex positions, in the latter stages of the simulation, once the system energy was $< 50\%$ of the possible minimum, the triskelion centres were allowed to move freely.

Our triskelion model is unchanged by the application of an arbitrary rotation to all vectors associated with a bead type: since the internal interactions between different beads fix their relative orientations, only the directions of the vectors for a particular type relative to each other are important. It is thus necessary to constrain some vectors during interaction-finding simulations. Without loss of freedom, for each bead type $B - E$, we chose to fix the bending vector used for the interaction with the previous bead in the leg. Additionally, the internal torsional vector for type $A$, which lies on its axis of symmetry, was fixed, and the three bending vectors of $A$ for interactions with $B$ beads were constrained to not rotate around the symmetry axis. The internal interaction vectors for bead type $A$ were such that there was a threefold rotational symmetry axis through the centre. 

If such constraints were not applied, vectors would be able to explore all directions and their average would be become undefined. Furthermore, even with the constraints already outlined, there is a large degeneracy in possible torsional vectors, both for internal and external interactions. Therefore, internal torsional vectors, for all bead types except $A$, were adjusted for each configuration to make them as close to perpendicular to the bending vectors as possible, without altering the resulting triskelion. Furthermore, the external torsional vector for bead type $B$ was taken to be $\mathbf{v}^{*}[B,2] \times \mathbf{v}^{*}[B,1] / \left|\mathbf{v}^{*}[B,2] \times \mathbf{v}^{*}[B,1]\right|$. For the other external torsional vectors, a minimum angle between attractive patch vectors and their corresponding torsional vector was imposed.

Not all interaction-finding simulations were found to converge to a low-energy minimum with all patches interacting as desired. Therefore, an intermediate configuration from a successful simulation, for which all patches were interacting, was chosen as a starting point for further simulations. These were run with 3 repeats each at $k_BT = 10^{-2}, 10^{-3}$ and $10^{-4} \epsilon_{tt}$. At a given $k_BT$, all simulations were found to converge to states fluctuating around the same energy value. As expected, these energy values, and the size of the fluctuations, both relative to $\epsilon_{tt}$, decreased with decreasing $k_BT$. All were within a range of $\approx \epsilon_{tt}$ around $-310 \epsilon_{tt}$, compared to a possible minimum of $-324 \epsilon_{tt}$ if each attractive patch on each triskelion interacted with one other patch and all these interactions were minimised. 

Sets of parameters for free assembly simulations were extracted as thermal averages, taken after the simulation had relaxed to the minimum, $d = \left<d^{*}\right>$, $\left\{\mathbf{v}[\alpha,i]\right\} = \left\{\left<\mathbf{v}^{*}[\alpha,i]\right>/\left|\left<\mathbf{v}^{*}[\alpha,i]\right>\right|\right\}$ and $\left\{\mathbf{u}[\alpha,i]\right\} = \left\{\left<\mathbf{u}^{*}[\alpha,i]\right>/\left|\left<\mathbf{u}^{*}[\alpha,i]\right>\right|\right\}$. The parameters from all simulations at different $k_BT$ were very similar. To test the size of the differences between the different parameter sets,  triskelia defined by each were placed, in their mechanical equilibrium configurations, with the same centre position and with the centre symmetry vector and the direction of the first leg aligned. The maximum distance between corresponding beads in different triskelia was $7\times10^{-3}\sigma$ and the largest angle between patches was $4\pi\times10^{-3}$. The largest angle between torsion vectors was larger, $6\pi \times 10^{-2}$, due to the larger freedom in choosing these. Given the close similarity, we simply picked one set and verified that these parameters did produce triskelia that can form a hexagonal barrel with energy $\approx -310 \epsilon_{tt}$ for low $k_BT$. 

We give the parameters used in our free assembly simulations. The equilibrium separation between internally bonded triskelion beads was $d = 0.81597\sigma$, to 5 decimal places. The vector parameters used for defining internal, $\left\{\mathbf{u}[\alpha,i]\right\}$,  and external, $\left\{\mathbf{v}[\alpha,i]\right\}$, interactions are given in Tables~\ref{tab:u_vecs} and~\ref{tab:v_vecs} respectively.

\subsection{\label{app:mem} Membrane Model}

We next discuss our membrane model. For the interactions between membrane particles, we use smooth potentials that are appropriate for molecular dynamics~\cite{noguchi2005}. Bonded membrane particles interact via
\begin{equation}
U_{bond}(r_{ij}) = 
\left\{
\begin{array}{l}
0\\
\hspace{1.2 in} \mathrm{for} \; r_{ij} \le 1.15\sigma, \\
(80 k_BT)\exp[1/(1.15\sigma - r_{ij})] / (1.33\sigma - r_{ij})\\
\\
\hspace{1.2 in} \mathrm{for} \;  1.15\sigma < r_{ij}  < 1.33\sigma, \\
\infty \\
\hspace{1.2 in} \mathrm{for} \; r_{ij} \ge 1.33\sigma, \\
\end{array}
\right.
\end{equation}
with $r_{ij} = |\mathbf{r}_{ij} | =  | \mathbf{r}_j - \mathbf{r}_i | $, where $\mathbf{r}_{i}$  is position of particle $i$. All pairs of membrane particles experience an excluded volume potential
\begin{equation}
U_{EV}(r_{ij}) = 
\left\{
\begin{array}{l}
\infty \\
\hspace{1.2 in} \mathrm{for} \; r_{ij} \le 0.67\sigma, \\
(80 k_BT)\exp[1/(r_{ij} - 0.85\sigma)] / (r_{ij} - 0.67\sigma)\\
\\
\hspace{1.2 in} \mathrm{for} \; 0.67\sigma < r_{ij} < 0.85\sigma,\\
0\\
\hspace{1.2 in} \mathrm{for} \; r_{ij} \ge 0.85\sigma.\\
\end{array}
\label{eq:mem_pot}
\right.
\end{equation}
The minimum distance between any two membrane particles is $0.67\sigma$ and the maximum bond length is $1.33\sigma$. 

Since we employ an SRD solvent, which requires the simulation box to be divisible into a regular grid of cells, the approach of using box-rescaling for simulating a tensionless membrane from our previous work without solvent~\cite{matthews2012} is not suitable. We therefore simulate the membrane by restricting a number, $N_{border}$, of particles, denoted border particles, to a frame with a confining potential. These form the edge of the membrane. Other, bulk, particles do not experience the confining potential. Furthermore, bonds between membrane particles are also of border and bulk types, where the border bonds are always between two border particles and always form a single closed ring. The bonding potential for both bond types is the same. 

\begin{figure}[ht!]
\begin{center}

\includegraphics[scale=0.3]{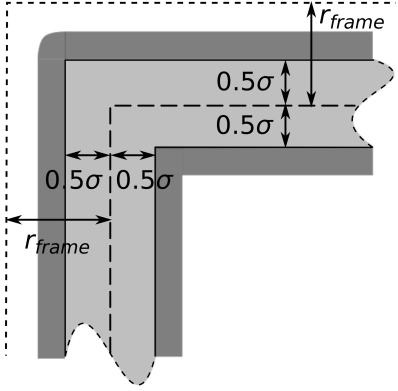}

\caption{\label{fig:r_frame} Sketch of a section of the frame in a simulation box, showing a cut in the plane of the frame. The outer, short-dashed line shows the periodic boundaries of the simulation box. The centre of the frame, shown by the long-dashed line is located a distance $r_{frame}$ inwards from this. There is a region where border membrane particles experience a flat potential, extending $0.5\sigma$ in either direction from this, shown by the light grey area. In the dark grey area, the border membrane particles experience a confining potential that diverges at the edge further from the flat region. In the direction out of the plane (not shown) the flat-potential region extends $4\sigma$.}
\end{center}
\end{figure}

The position of the frame is given by a variable $r_{frame}$ that defines how far in from the edges of the simulation box it is, see Fig.~\ref{fig:r_frame}. In the direction out of the plane of the frame a flat-potential region extends for $4\sigma$. In the plane it extends $0.5\sigma$ inwards and outwards from the frame position, see Fig.~\ref{fig:r_frame}. 

In the flat-potential region, border membrane particles have an energy of $E_{frame}$. The confining potential for the border membrane particles is of the same form, range and strength as the excluded volume between membrane particles,
\begin{equation}
U_{confine}(r) = 
\left\{
\begin{array}{l}
E_{frame} + (80 k_BT)\exp[-1/r] / (0.18\sigma - r)\\
\\
\hspace{1.2 in} \mathrm{for} \;  0 < r  < 0.18\sigma, \\
\infty \\
\hspace{1.2 in} \mathrm{for} \; r \ge 0.18\sigma, \\
\end{array}
\right.
\end{equation}
where $r$ is the distance of the border membrane particle from the closest point within the flat-potential region.

\begin{figure}[ht!]
\begin{center}

\includegraphics[scale=0.4]{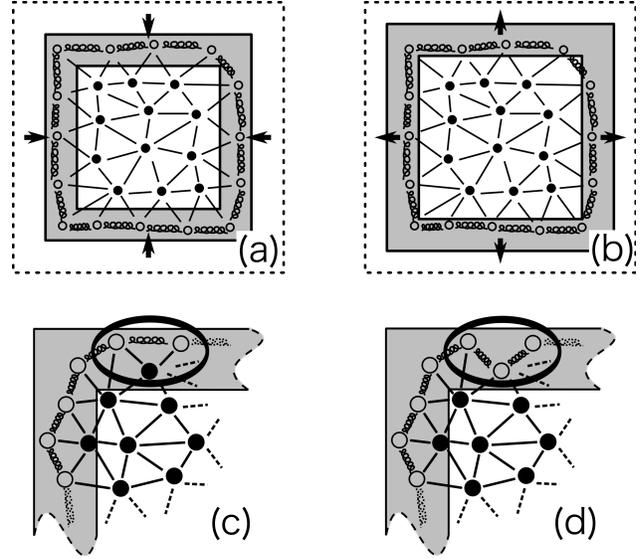}

\caption{\label{fig:membrane_border_2D} Sketch of the membrane boundary conditions used for dynamics simulations. Membrane particles are of two types: border (open circles) and bulk (filled circles), as are membrane bonds: border (loopy lines) and bulk (straight lines). In the shaded region, border particles experience a confining potential with a flat minimum region with energy $E_{frame}$, bulk particles feel no potential. Every so often, Monte Carlo updates of two types are performed: (a), (b) The position of the frame within the simulation box (dotted line) is shifted inwards or outwards. (c), (d) Bulk particles are converted to border particles, see highlighted area, or vice versa. Only bulk particles that are bonded to two current border particles, which are themselves bonded, may be converted. A corresponding bond creation (deletion), along with a conversion of existing bonds from bulk to border (border to bulk) ensures that each border particle has two, and only two, border bonds to two other border particles, and that the bonds between all the border particles form a closed ring. A double bond between the same two particles may not be created and a border bond may only be deleted if the particle that thus becomes part of the border was originally a bulk particle.}
\end{center}
\end{figure}

As summarised in Fig.~\ref{fig:membrane_border_2D}, we employ two types of MC moves to allow the membrane to expand and contract: those that convert border particles to bulk particle and vice versa, as well as those that change $r_{frame}$, moving the frame inwards or outwards from the simulation box centre. The total number of bonds and triangles, $N_{tri}$, in the membrane may thus vary. Maximum and minimum values of $r_{frame}$ are imposed such that the confining potentials from opposite sides of the frame do not overlap, and also such that the confining potential does not extend across the periodic boundaries.

The value of $E_{frame}$ controls $r_{frame}$ and $N_{border}$: if it is made very high the membrane will try to minimise $N_{border}$, and consequently $r_{frame}$ will increase as the membrane is squeezed, forcing it to extend out of the plane. On the other hand, if $E_{frame}$ is made very low, $N_{border}$ will become large and $r_{frame}$ small, and the central region of the membrane will  be stretched into a flat configuration. To determine $E_{frame}$ for a particular membrane stiffness, we simulated membranes with a range of $E_{frame}$ and compared the ratio of the area to the area projected onto the plane of the frame, $A/A_{proj}$, to the value measured for a tensionless membrane simulated with box-rescaling~\cite{matthews2012}. To minimise the effect of the frame, we only considered the part of the membrane with a projection falling within the central $10\%$ of the area defined by the frame. 

A unit normal vector is associated with each membrane triangle. Each bulk bond forms the side of two different neighboring triangles. Membrane fluidity is included using MC moves that attempt to remove a given bulk bond and create a new one between the two vertices of its neighboring triangles that were not connected by the original. During dynamical simulations,  a number of moves equal to the number of bulk bonds are performed every $0.1 t_0$. By simulating a Poiseuille flow for a two-dimensional membrane~\cite{noguchi2005}, we estimate the resulting viscosity of the membrane to be $35.1 \pm 0.1m/ t_0$. During the bond-flipping procedure, the direction of the normals is always maintained such that, if the membrane were in a flat configuration, all normals would be aligned. 

The bending stiffness of the membrane is controlled by including a potential $U_{bend} = \lambda_b (1 - \mathbf{n}_i \cdot \mathbf{n}_j)$ for each bond, where $ \mathbf{n}_i$ and $ \mathbf{n}_j$ are the unit normal vectors of the two triangles neighboring the bond and $\lambda_b$ is an energy. The total membrane area, $A$, is constrained with a harmonic potential, $U_{area} = (k_BT)(A - A_0)^2$, where $A_0 = (\sqrt{3}/4) l^2 N_{tri}$, in the Hamiltonian. Additionally, a bending potential of the same form is applied to triangles with a border bond as one of their edges, where the unit normal of the triangle is compared to a unit normal to the frame-plane.

In our dynamical simulations, a series of MC moves, each changing $r_{frame}$ by  $\Delta r_{frame}$  or $N_{border}$ by $\pm 1$, were performed every $0.1 t_0$. On average there were $10^3$ attempted changes to $r_{frame}$ and $10 N_{mem}$ attempted moves to change $N_{border}$. The acceptance ratio for the $r_{frame}$-moves showed some dependence on the simulation parameters and also changed somewhat during the course of a simulation, for example as a bud was formed. It was however found to always be roughly in the range $0.3 - 0.4$. The acceptance ratio for the $N_{border}$-moves depended on the membrane stiffness, ranging from $\approx 5 \times 10^{-2}$ for $\lambda_b = 8\sqrt{3} k_BT$ to $\approx 7 \times 10^{-2}$ for $\lambda_b = 2\sqrt{3} k_BT$. A similar, though weaker, dependence was seen for the acceptance ratio of bond flips, which had a value of $\approx 2 \times 10^{-2}$, similar to that seen in previous work with the same membrane model~\cite{noguchi2005}.

The rate of $N_{border}$-moves chosen means there were $\approx 600$ such moves accepted every $0.1 t_0$. Similarly, there were $\approx 300$ $r_{frame}$-moves accepted. $\Delta r_{frame}$ values are chosen uniformly in the range $-0.05\sigma < \Delta r_{frame} < 0.05 \sigma$ so if these $\approx 300$ shifts were an unbiased random walk, $r_{frame}$ would explore a range of $\approx 0.2 \sigma$. Given that the values of $N_{border}$ and $r_{frame}$ changed by at most $\approx 80$ and $\approx 7\sigma$ respectively over the course of a $2.5 \times 10^4 t_0$ simulation, the rate of $N_{border}$- and $r_{frame}$-moves are sufficiently high that they will adjust the system to the local minimum every $0.1 t_0$ and their exact values will not affect the results.

We next discuss the triskelion-membrane particle interactions, the form of which is given in Eq.~\ref{eq:LJ_pot}. For these interactions, the energy scale, $\epsilon$, is set to $\epsilon_{mt}$. Since the membrane bonds have a relatively broad, flat minimum, the membrane particles would tend to be locally compressed when an attractive triskelion bead is close. For $mt$ interactions the $\gamma_{area}$ factor is used to counter-act this by making the interaction proportional to the area that the membrane particle represents: $\gamma_{area} = A_{neigh}/(N_{neigh}A_{tri}$), where $N_{neigh}$ is the total number of triangles that have the membrane particle as a vertex, $A_{neigh}$ is their total area and $A_{tri} = A_0 / N_{tri}$. 

For $mt$ interactions, $\gamma_{att} = 0$ for triskelion bead types $A - D$. For bead type $E$ it is used to make only one side of the membrane attract the triskelion beads: it takes a value of $1$ if the bead is ``above'' the membrane and $0$ if it is ``below''. A triskelion bead is determined to be ``above'' or ``below'' by finding the closest point  on the membrane. If the normal of the triangle enclosing the closest point makes an angle of less than $\pi / 2 $ with the vector from the closest point to bead then the bead is ``above'' the membrane, otherwise it is ``below''. For $mt$ interactions $\gamma_{orient} = 1$ since these are chosen not to be patchy. 

Since the attractive interaction between triskelia and the membrane depends on which side of the membrane they interact with, a discontinuity in the potential would arise if triskelia could move from one side of the membrane to the other whilst remaining within the interaction range. To avoid this, triskelia experience an excluded volume potential around the frame. So that the total assembly volume available to the triskelia remains approximately constant, the width of the excluded volume region around the frame is rescaled as it moves so its volume is not changed. This excluded region does not generally extended across the entire simulation box.

\subsection{\label{app:SRD} SRD}

We finally discuss our choice of SRD parameters. The side of the SRD cells was chosen to be equal to $\sigma$. Both membrane particles and the beads forming triskelia were given masses of $5m$ and coupled to the solvent via the collision step~\cite{gompper2009}. The rotational degrees of freedom of the beads are thus not coupled directly to the solvent but, given the relatively stiff triskelia we simulate, these relax rapidly anyway. In mechanical equilibrium the separation between the central triskelion bead and the final bead of a leg is about $2.8\sigma$  and that between the final beads of different legs is about $3.4\sigma$. A triskelion thus spans many SRD cells and so its rotation as a whole is coupled to the solvent. We chose to have a number density of 5 SRD particles per cell and performed collisions every $0.1t_0$ with a rotation angle of $\frac{\pi}{2}$, giving a viscosity of $\eta = 2.5 m/\sigma t_0 $~\cite{kikuchi2003}. We apply a momentum-conserving cell-level thermostat to the SRD fluid~\cite{gompper2009}.



\footnotesize{
\providecommand*{\mcitethebibliography}{\thebibliography}
\csname @ifundefined\endcsname{endmcitethebibliography}
{\let\endmcitethebibliography\endthebibliography}{}

}

\end{document}